\documentclass[12pt]{article}
\pdfoutput=1
\usepackage{jheppub}
\usepackage[skip=2pt,font=footnotesize]{caption}
\usepackage[utf8]{inputenc}
\usepackage{verbatim}
\usepackage{amsmath}
\usepackage{amssymb}
\usepackage{amsthm}
\usepackage{slashed}
\usepackage{amsfonts}
\usepackage{makecell}
\usepackage{tabularx}
\usepackage{comment}
\usepackage{subfigure}
\usepackage{float}
\usepackage{color}
\usepackage{physics}

\newtheorem{theorem}{Theorem}
\newtheorem{definition}[theorem]{Definition}

\newtheorem{conj}{Conjecture}

\usepackage{slashed}
\renewcommand{\d}{\partial}
\newcommand{\dmax}{\Delta_{\rm max}}
\newcommand{\half}{\frac{1}{2}}
\newcommand{\CO}{\mathcal{O}}

\newcommand{\CM}{\mathcal{M}}

\renewcommand{\>}{\rangle}
\newcommand{\<}{\langle}
\newcommand{\pr}[1]{\left( #1\right)}
\newcommand{\corr}[1]{\left\langle #1\right\rangle}

\begin{document}


\title{Quantum simulation of quantum field theories as quantum chemistry}

\author[a,b]{Junyu Liu}
\author[c,d]{Yuan Xin}

\affiliation[a]{Walter Burke Institute for Theoretical Physics,\\ California Institute of Technology, Pasadena, CA 91125, USA}
\affiliation[b]{Institute for Quantum Information and Matter,\\ California Institute of Technology, Pasadena, CA 91125, USA}
\affiliation[c]{Department of Physics, Boston University,\\ 590 Commonwealth Avenue, Boston, MA 02215, USA}
\affiliation[d]{Department of Physics, Yale University,\\ New Haven, CT 06520, USA}
\emailAdd{jliu2@caltech.edu}
\emailAdd{yuan2015@bu.edu}

\abstract{Conformal truncation is a powerful numerical method for solving generic strongly- coupled quantum field theories based on purely field-theoretic technics without introducing lattice regularization. We discuss possible speedups for performing those computations using quantum devices, with the help of near-term and future quantum algorithms. We show that this construction is very similar to quantum simulation problems appearing in quantum chemistry (which are widely investigated in quantum information science), and the renormalization group theory provides a field theory interpretation of conformal truncation simulation. Taking two-dimensional Quantum Chromodynamics (QCD) as an example, we give various explicit calculations of variational and digital quantum simulations in the level of theories, classical trials, or quantum simulators from IBM, including adiabatic state preparation, variational quantum eigensolver, imaginary time evolution, and quantum Lanczos algorithm. Our work shows that quantum computation could not only help us understand fundamental physics in the lattice approximation, but also simulate quantum field theory methods directly, which are widely used in particle and nuclear physics, sharpening the statement of the quantum Church-Turing Thesis. }

\maketitle

\section{Introduction}
Quantum field theory is one of the most astonishing tools we have for the understanding of the universe. Physicists use quantum field theory to make predictions from fundamental particles, early universe, to complex behaviors in condensed matter systems. When comparing with experiments, physicists could use their pens to perform computations following rules of quantum field theory, or if in a more complicated situation, use their computers to make numerical predictions.  

Although there is no rigorous proof, people usually tend to believe that the universe is simulatable by their most powerful computers. A more formal statement is so-called the quantum Church-Turing Thesis\footnote{There are several variants of the quantum Church-Turing Thesis. A particular important variant is based on the complexity case, emphasizing the efficiency of simulations, which is called quantum \emph{complexity-theoretic} Church-Turing Thesis or quantum \emph{extended} Church-Turing Thesis. Here we just call all of them quantum Church-Turing Thesis in general.}, 
\begin{conj}
If there is a physical system that could naturally perform a calculation consistent with the law of the universe, then this calculation could also be performed by a quantum Turing machine. 
\end{conj}

However, there are some negative results towards the above conjecture. In fact, some problems with many-body Hamiltonians are shown to be QMA-complete, a complexity class that is hard even for quantum computers in the worst case \cite{liu2006consistency,liu2007quantum,kempe20033,kempe2006complexity}. Quantum field theory, defined on continuous spacetime with infinite sites, where each site supports an infinite-dimensional Hilbert space, seems to be much harder than many-body physics (although there is a discussion of BQP-completeness in quantum field theory \cite{jordan2018bqp}). Its complication especially emerges at strong coupling, where we usually do not have sufficient analytic understandings with precise Lagrange descriptions. For strongly coupled field theory existing in the real world, for instance, the strongly-coupled Quantum Chromodynamics (QCD), people usually discretize the space on the lattice and try to simulate theories with Lagrange description starting at weak coupling, which is an active area still in development. Performing high-quality lattice gauge theory calculation requires huge computational resources and sometimes intractable due to large Hilbert space dimensions and the sign problem. Recently, motivated by rapid development in quantum information science, people tend to study potential quantum speedup for lattice gauge theory or quantum field theory in general or try to put latticed quantum field theory calculation in the real hardware (for instance, see a collection of references \cite{jordan2016black,Jordan:2011ne,Jordan:2011ci,jordan2014quantum,jordan2017fast,moosavian2018faster,moosavian2019site,klco2018quantum,Klco:2018zqz,Lu:2018pjk,Klco:2019xro,Klco:2019evd,Klco:2019yrb,Klco:2020aud,Alexeev:2019enj,Preskill:2018fag,huang2020predicting,Hu:2018hyd,Zou:2019xbi,hauke2012speeding,barrett2013simulating,zohar2013quantum,zohar2013cold,perez2014mapping,brennen2015multiscale,wiese2014towards,garcia2015fermion,marcos2014two,wiese2013ultracold,zohar2015quantum,martinez2016real,bermudez2017quantum,gonzalez2017quantum,muschik2017u,li2018digital,meurice2018tensorial,raychowdhury2018tailoring,roy2019quantum,mishra2019quantum,singh2019qubit,banuls2019simulating,mueller2019deeply,bao2019quantum,bauer2019quantum,lamm2019general,yeter2019scalar,kharzeev2020real,farrelly2020discretizing,alexandru2020quantum,vovrosh2020confinement,kreshchuk2020quantum,chakraborty2020digital,harmalkar2020quantum}). 

Nevertheless, it motivates us to ask the following question: can we get rid of lattice and study quantum field theory directly, in order to avoid large Hilbert spaces localized on each site? In fact, it might be possible. Since we are mostly interested in the strong coupling, physics at the UV fixed-point is described by conformal field theory (CFT), whose predictions are analytically tractable in part. For some CFTs, we know analytically, or numerically, their spectra and strengths of three-point functions (the Operator Product Expansion, OPE coefficients). If we study a generic quantum field theory at strong coupling, we could turn on some relevant operators away from the UV fixed point. Thus, taking advantage of known conformal data, we could form a Hamiltonian for such a generic quantum field theory: its diagonal terms are given by the energy spectrum of the CFT, while its off-diagonal terms are given by OPE coefficients. A proper cutoff of the dimension of the Hilbert space will make the approximate Hamiltonian finite-dimensional\footnote{It makes sense to introduce a cutoff.  Generic CFTs will behave like mean-field theories at large spin. Thus, OPE coefficients will quickly decay at high spin \cite{Komargodski:2012ek,Alday:2015eya}.}. Thus, based on the above treatment, we reduce the simulation task of the corresponding quantum field theory as diagonalizing and evolving a possibly large matrix. This method is called conformal truncation or Hamiltonian truncation in quantum field theory community (see some recent discussions, for instance, \cite{Hogervorst:2014rta,Rychkov:2014eea,Rychkov:2015vap,Katz:2013qua,Elias-Miro:2015bqk})\footnote{There exists a closely-related approach, which is called light-front quantum field theory. This terminology is usually used for high energy phenomenology or high energy lattice community, specifically designed for strongly coupled QCD and hadronic physics. See some recent reviews for instance, \cite{Bakker:2013cea}. }. 

Thus, other than considering a quantum many-body problem with local interaction in a lattice, now we formulate a quantum-mechanical problem with an explicit Hamiltonian, without any notion of locality. Although it is not clear if this treatment is generically better than lattice regularization, we will hope that it will open up different possibilities for quantum field theory simulation at strong coupling. The computational task of conformal truncation is also very hard, requiring significant computational resources, especially for field theory in higher dimensions. Thus, conformal truncation provides another opportunity for quantum devices. In fact, generic quantum-mechanical problems without locality are widely investigated already in the quantum information community: quantum chemistry (See a recent review \cite{mcardle2020quantum} and Table \ref{table:1}). 

\begin{table}[h!]
\centering
\begin{tabular}{|c c c|} 
 \hline
 Treatment & Starting point & Existing toolbox in QI  \\ [0.5ex] 
 \hline\hline
 Lattice regularization & Usually free theory & Quantum many-body physics \\
 Conformal truncation & Strongly-coupled theory (CFT) & Quantum chemistry \\
 \hline
\end{tabular}
\caption{Simulating quantum field theories using different quantum information (QI) methods. We compare the lattice method and the conformal truncation method and point out potential tools we need to use from existing methods in quantum information science. For simulating quantum field theories with lattice, we could use similar methods developed for quantum many-body physics. While without lattice, we could use methods in quantum chemistry to simulate conformal truncation.}
\label{table:1}
\end{table}

We might also ask the following question. Why do we need quantum computing? We could summarize the reasons as the following. 
\begin{itemize}
\item First, the conformal truncation program meets real challenges about computational powers. Even if starting from 
mean-field
theories in UV CFT, the conformal truncation program is very hard to perform at, especially, higher-dimensional field theories, for instance, 2+1 dimensional or higher. This is because the required matrices we have is too large. Thus, quantum computing might help potentially for studying their spectra or real-time dynamics. 
\item Second, the study of quantum simulation for conformal truncation should be helpful for the claim of quantum supremacy. Imagine that we have a quantum simulation task in latticed quantum field theories, and we show that quantum computing is extremely hard in such a task. If one could show that the conformal truncation problem is more suitable for quantum computers in this situation, it might be easier to arrive at quantum supremacy. 
\item Finally, the study of quantum simulation for conformal truncation problems is helpful for conceptual understanding of the quantum Church-Turing Thesis. As we discussed before, conformal truncation might provide alternative aspects for quantum simulation of quantum field theories, which might support or weaken the statement of the quantum Church-Turing Thesis.
\end{itemize}
In this paper, we will show how to use methods from quantum chemistry to simulate conformal truncation. The work is organized as the following.
\begin{itemize}
\item In Section \ref{theory}, we discuss our theoretical framework for conformal truncation and quantum simulation interpretation of it. This will involve an introduction for conformal truncation, a discussion about the renormalization group in digital or variational quantum simulation in the context of conformal truncation, and a study about generic features of conformal truncation problems that might be helpful for quantum simulation.   
\item In Section \ref{digital}, we introduce digital quantum simulation for conformal truncation problems. Our simulation for low-lying states is based on adiabatic state preparation (ASP) and oracle-based algorithms. These simulation methods could be applied to a potential universal quantum computer.  
\item In Section \ref{QCD}, we give an introduction to our primary example, the two-dimensional quantum chromodynamic (QCD). 
\item In Section \ref{variational}, we introduce variational quantum simulation for conformal truncation. We mostly use variational quantum eigensolver (VQE) for solving low-lying excited states. We give examples in theory, classical trial numerics, and quantum simulation numerical experiments in \texttt{IBM quantum experience}. We also introduce a short discussion about encoding.
\item In Section \ref{other}, we introduce other near-term algorithms. This section involves a discussion about imaginary time evolution (ITE) and classical or quantum versions of the Lanczos algorithm. We also develop a novel variational-based quantum Lanczos algorithm. We provide trial numerics to justify the usage of the above algorithms. 
\item In Section \ref{conc}, we give a conclusion and overview of potential research direction following the line of this work. 
\item In Appendix \ref{technical}, we give some technical computations on 2D QCD.
\end{itemize}

Studying quantum simulation of quantum field theory will also be helpful for other quantum physics. For instance, in the recent studies of quantum gravity and quantum black holes, people use quantum-mechanical models without locality to mimic maximal chaotic behaviors of quantum black holes, for instance, the Sachdev-Ye-Kitaev (SYK) model \cite{Sachdev:1992fk,kitaev,Maldacena:2016hyu}. Currently, efforts have been made for making SYK-type models into digital or analog platforms \cite{Garcia-Alvarez:2016wem,Danshita:2016xbo,Brown:2019hmk,Liu:2020sqb}. Our work might be helpful for simulations of those models. Moreover, it provides a novel method to simulate quantum field theories beyond the usual lattice regularization. Combining other efforts for making quantum field theory computations suitable for quantum devices (for instance, see \cite{bao2019quantum}), it might be helpful for novel and technical computations for strongly-coupled quantum field theories, and benchmarks for near-term quantum devices. Some more details are summarized in Section \ref{conc}.

Finally we wish to mention that there is a related nice paper \cite{kreshchuk2020quantum} about quantum simulation for light-front physics appearing earlier this year. Part of this work about digital algorithms is partially inspired by \cite{kreshchuk2020quantum}. In terms of 2D QCD, the primary example that is considered in this work, the conformal truncation method is significantly different from light-front treatment by the following,
\begin{itemize}
\item We are using the conformal basis. This basis forms a set of eigenstates in the UV CFT. Thus, we have a clear interpretation of the renormalization group theory. The basis used in \cite{kreshchuk2020quantum}, since constructed from different motivations, does not have such an interpretation. 
\item The aim of simulating light-front physics is mostly from high energy lattice and high energy phenomenology. Thus for light-front physics, people will mainly look at physical theories such as Yukawa model or QCD. However, the scope of this work will be more general strongly-coupled theories, although we take 2D QCD as our simplest example. 
\item The basis used in \cite{kreshchuk2020quantum} (and other typical bases in light-front physics), is constructed by compactifying the space in a finite regime and imposing some boundary conditions. Here, we directly treat infinite space, and our momenta are continuous.  
\end{itemize}
It is not clear which approach is better, and it highly depends on the motivation of simulation. Although the motivation and treatment of the work \cite{kreshchuk2020quantum} are very different from us, some technical approaches might be similar. For instance, the encoding method and the quantum simulation algorithms might be similar,  inherited from the quantum simulation of quantum chemistry. We look forward to furthering research on comparing those methods.

\section{Quantum field theory without lattice}\label{theory}
\subsection{Introduction to conformal truncation}
\label{sec:introToConformalTruncation}

Here we use $D$ to denote the spacetime dimension. For 2D QCD example we take $D=2$.

As we have explained in the introduction, quantum field theory is ubiquitous but is notoriously hard at strong coupling. The frequent strategy of studying quantum field theories regularized by a lattice has its limitations. A different approach, known as the {\it conformal truncation} method, utilizes Hamiltonian truncation\footnote{In this paper, sometimes we abuse the terminology of \emph{Hamiltonian truncation}. Sometimes we mean truncating the Hamiltonian for generic quantum mechanics, while sometimes we mean truncating the Hamiltonian for quantum field theories. We believe there is no confusion based on the environment of the text.} and CFTs. The Hamiltonian truncation, also known as the {\it Rayleigh-Ritz} method, is a variational method of finding the approximate spectrum and eigenstates of a Hamiltonian using a finite number $n_{\rm max}$ of basis states. A typical strategy is to choose the basis to be the eigenstates of an exactly-known part $H_0$ of the Hamiltonian, and study the full Hamiltonian $H = H_0 + g V$, with respect to the basis up to the truncation $n_{\rm max}$. 

To study quantum field theory using the Hamiltonian truncation method, one needs a theory that reliably provides $H_0$ and the basis states, and additionally, one needs to discretize the continuous Hilbert space in order to keep a finite number of representative basis states. The conformal truncation, first pioneered by Yurov and Zamolodchikov in the work 
\cite{yurov1990truncated} known as Truncated Conformal Space Approach (TCSA), views the Hamiltonian as a CFT deformed by a relevant operator 
\begin{equation}
\label{eq:hamiltonian}
H = H^{\rm (CFT)} + g \int d^{D-1} x \, \CO_R (x)~,
\end{equation}
where the coupling $g$ can be large.

The motivation of using CFT is both philosophical and practical. 
\begin{itemize}
\item Philosophically, conformal truncation originates naturally from the renormalization group (RG) flow. The quantum field theory is defined as RG flow between short distance (UV) and long-distance (IR) fixed points. The dimension of $\CO_R$ is less than $D$, so $H^{\rm (CFT)}$ describes the UV fixed point and the RG flow will take the quantum field theory away from the UV CFT and take it to a new IR fixed point, which can be another CFT or a theory with a mass gap.
\item Practically, one can organize the basis according to the conformal symmetry and compute the Hamiltonian using the highly constrained conformal algebra. 
\end{itemize}

In this work, we study a setup known as {\it Lightcone Conformal Truncation} 
\cite{pedagogical,Katz:2013qua,Katz:2014uoa,Katz:2016hxp,Anand:2017yij,Fitzpatrick:2018ttk,Delacretaz:2018xbn,Fitzpatrick:2018xlz,Fitzpatrick:2019cif,Anand:2019lkt,2dQCDToAppear}, where we use lightcone quantization to compute conformal truncation in infinite volume. In the traditional setups of conformal truncation, one needs to put the theory in either a finite-sized box or a compact sphere in order to discretize the Hilbert space and extrapolate the result to infinite volume. The extrapolation suffers from a conceptual issue known as the {\it orthogonality catastrophe}\footnote{Our discussion of orthogonality catastrophe in the context of conformal truncation follows \cite{pedagogical}.} \cite{anderson1967infrared}.

In an interacting theory, the UV vacuum $\ket{0}$ generically gets renormalized, and the new interacting vacuum $\ket{\Omega}$ involves more and more states as the volume grows. A simple dimensional analysis shows that since the entropy grows as the volume, the overlap between the physical vacuum and the UV vacuum is exponentially suppressed
\begin{equation}
|\< 0 | \Omega \>|^2 \sim e^{-S }~, \qquad S \sim L^{D-1} ~,
\end{equation}
which means with growing volume, it will be exponentially hard to find the interacting vacuum $| \Omega \>$ starting from the free vacuum $| 0 \>$. In lightcone quantization, we directly study infinite volume flat space, with the coordinate system
\begin{equation}
x^\pm \equiv \frac{1}{\sqrt 2} (t \pm x^1)~, \qquad 
\mathbf{x}^\perp \equiv (x^2, x^3, \cdots, x^{D-1} )~,
\end{equation}
where we take the \emph{lightcone time} to be $x^+$. The \emph{lightcone Hamiltonian}, $P_+$, is the generator of the $x^+$ translation. The new mass shell relation is
\begin{equation}
m^2 = 2p_- p_+ - |\mathbf{p}_{\perp}|^2 ~,
\end{equation}
where Lorentz boost in the $x_1$ direction is trivialized: it simply rescales $p_-$ and $p_+$ keeping the product, and one can always set $p_- = 1$ using Lorentz invariance, and $p_+$ is linear in the mass or center-of-energy $m^2$. Note that this relation separates the vacuum, whose $p_- = 0$, from the rest of the states\footnote{Strictly speaking, the vacuum is not the only state with $p_- = 0$, there is a measure-zero set of states called the {\it zero mode}. The zero-mode can be integrated out, giving a correction to the Hamiltonian, and the conclusion that the lightcone vacuum does not renormalize is still valid. See \cite{Fitzpatrick:2018ttk} for details.}. As a result, the vacuum does not renormalize, and the infinite volume limit does not suffer from the orthogonality catastrophe.

In LCT we construct the infinite-volume conformal basis by acting the primary operators $\CO(x)$ in the UV CFT on the vacuum, and Fourier transform as the following
\begin{equation}
\label{eq:basis}
| \CO, p \> \equiv \int e^{i p x} dx \, \CO(x) | 0 \> ~.
\end{equation}
The primary operators are operators annihilated by the special conformal transition operator 
\begin{equation}
\label{eq:primCondition}
[K_- , \CO(0)] = 0 ~.
\end{equation}
Each state depends on the momentum $p$, the spin $\ell$, and the scaling dimension $\Delta$ of the operator $\CO$ and the representation of additional symmetry if applicable. In general, the basis is truncated to include a finite number of states up to a maximum conformal Casimir eigenvalue $\mathcal{C} \equiv \Delta  (\Delta -D)+\ell (D+\ell-2) \leq \mathcal{C}_{\rm max}$, and in the application of this work, 2D QCD, we can intuitively consider a truncation of maximum scaling dimension $\Delta \leq \dmax$. The state inner products and the Hamiltonian matrix elements can be computed from the CFT two-point functions
\begin{align}
\langle \CO, p | \CO^\prime, p^\prime \rangle & \equiv 
\int dx \, dy \, e^{i(px - p^\prime y)} \langle \CO(x) \CO^\prime(y) \rangle  \label{eq:InnerProduct} \nonumber \\ 
& = 
2p(2\pi) \delta(p-p^\prime) \, G_{\CO\CO^\prime}  ~,
\end{align}
and we normalize the states as $G_{\CO\CO^\prime} = \delta_{\CO\CO^\prime} $.

In LCT the Hamiltonian is the $P_+$ operator. As (\ref{eq:hamiltonian}) we take the UV CFT $H^{\rm (CFT)} \equiv P_+^{\rm (CFT)}$, which is exactly diagonalized in the basis
\begin{equation}
P_+^{\rm (CFT)} |\CO, p \> = p_+ | \CO, p \> ~,
\end{equation}
plus a deformation 
\begin{equation}
\delta P_+ \equiv g \int d^{D-1} x \, \CO_R (x) ~,
\end{equation}
and we obtain the full Hamiltonian $H \equiv P_+ = P_+^{\rm (CFT)} + \delta P_+$.

The matrix elements of the deformation can be computed using the CFT three-point correlation function
\begin{align}
\langle \CO,p | \delta P_+ |  \CO^\prime, p^\prime \rangle &\equiv 
\int dx \, dy \, dz\, e^{i(px - p^\prime z)} 
\langle \CO(x) \CO_R(y) \CO^\prime(z) \rangle 
\label{eq:MatrixElements}  \nonumber\\
& = 
2p (2\pi) \delta(p-p^\prime)  \CM_{\CO\CO^\prime}~.
\end{align}
Then, we could write down the finite-dimensional truncated Hamiltonian with respect to the finite basis truncated at $\dmax$. We diagonalize the truncated Hamiltonian and obtain the spectrum and eigenstates. This illustrates the general theoretical framework of conformal truncation. Later, we will give a more specific example, the 2D QCD to perform our simulations.
\subsection{Conformal truncation and quantum simulation}
\subsubsection{Quantum field theory without lattice}
As we have explained before, conformal truncation provides a natural formalism to approximate a quantum field theory in a finite-dimensional quantum mechanics. This finite-dimensional matrix does not have any manifest physical meaning of locality. At this stage, although people propose that conformal truncation might be a cheap alternative for lattice, it is not clear for us which method is better. A fair comparison with the same physical targets should be performed at some stage. But for us, since we make full usage of conformal data at the strong coupling, conformal truncation might be pretty natural to study static or real-time dynamics of quantum field theory. 

From a quantum simulation point of view, what we gain is a potentially more compact and possibly smaller Hilbert space comparing to lattice treatment, but we lose our control about the locality in such a Hamiltonian. This is physically natural since we start from the strong coupling. Thus, for a given Hamiltonian without locality, we should think about how to encode the Hamiltonian into a quantum computer. The situation is particularly similar to the study of quantum simulation in quantum chemistry. Because of its practical usage, quantum computing theories and algorithms are in rapid development (for instance, see \cite{arguello2019analogue}). Here, we will use \emph{quantum simulation without lattice} as a slogan and try to digest the recent development of quantum simulation for quantum chemistry for our strongly-coupled field theory problems. 

In terms of quantum simulation, conformal truncation also provides a natural physical interpretation in the language of RG flows. We will briefly describe possible explanations in the following two circumstances: digital and variational quantum simulation.

\subsubsection{Digital quantum simulation and RG flows}
As an example, in digital simulation for solving the ground state of the Hamiltonian, a trick we usually use is adiabatic state preparation. We start from a theory where the spectrum is known. We evolve the ground state of the theory with a time-dependent Hamiltonian. During evolution, we slowly turn on the coupling. Due to the adiabatic theorem, if the evolution is slow enough, the states in different time slices are always the ground states of the time-dependent Hamiltonian if there is no energy-level crossing. (We will describe this picture in more detail later). 

In the conformal truncation context, the starting Hamiltonian should be the UV CFT, and we finally wish to turn on the coupling, driving the theory away from UV fixed-point towards a general strongly-coupled quantum field theory. Thus, the path of adiabatic state preparation describes an RG flow, from the fixed point towards somewhere in the middle. This is a realization of an RG flow using quantum gates, where different steps of gates describe different theories. There is a pretty similar situation in high energy physics: sometimes we wish to change the energy scale of our theory to make predictions for different physics, and the coupling is dynamical in different energy scales. Thus, the adiabatic time evolution provides an analog of beta function in an RG flow.

\subsubsection{Variational quantum simulation and RG flows}
Here, we also present an interpretation of renormalization for the variational quantum simulation of conformal truncation. In conformal truncation, the full Hamiltonian with infinite entries is the honest theory in the strongly-coupled theory we want. When we truncate the spectrum, we obtain some theories on the more UV side. If we truncate the spectrum with only one element left, we get the ground state energy of the UV CFT. Thus, in this formalism, the RG flow is \emph{discrete}. Cutting off fewer elements means that we are able to probe a more accurate low energy theory. Thus, variational quantum simulation seeks accurate low-lying states for a given realization, is similar to the behavior of renormalization when we try to balance bared parameters and physical parameters to make predictions for a low energy effective field theory.

\subsubsection{Features of conformal truncation problems and a natural variational ansatz}
Here we briefly describe the advantage of the conformal basis in conformal truncation and generic features of sparsity in conformal truncation matrices. 

The conformal truncation is, like other Hamiltonian methods, naturally a variational task, where the quality of the result is determined by the basis states. There are apparent gains from the conformal basis. First, the states directly describe the physical degrees of freedom of a quantum field theory. This is, in contrast to, studying quantum field theories as the continuous limit of a lattice, where the on-site degrees of freedom are not directly related to a propagating field. As a result, conformal truncation usually approaches the result of similar quality with a much smaller Hilbert space. Second, the conformal basis has more spacetime symmetries built-in. More symmetry means simpler Hamiltonian, less divergence, and less fine-tuning. On a lattice, the rotation and translation symmetry are tricky, and it is not at all clear how to preserve Lorentz invariance. For instance, using the LCT framework, one preserves all the above symmetries, and the work \cite{Fitzpatrick:2019cif} shows supersymmetry can also be preserved without fine-tuning. Finally, conformal truncation, by definition, gives the RG flow a stable UV fixed point, and the quantum field theory is well-defined at each point on the RG flow from the UV CFT to the IR fixed point. It is possible to study the real-time dynamics at all energy scales, not just the ground state. 

The conformal truncation problem has its own properties. Lattice Hamiltonian is local and sparse because the local interaction between lattice sites gives the banded structure of the Hamiltonian matrix, and the number of nonzero matrix elements grows slower than the size of the matrix because the interaction shuts off beyond the interaction range, giving a sparse matrix. In LCT, the states are organized by their scaling dimension $\Delta$ instead of lattice site, and the matrix elements between a pair of states are suppressed by the difference of scaling dimensions $\Delta_i - \Delta_j$,  and the off-diagonal matrix elements are punished. The suppression goes as a power law of $\Delta_i - \Delta_j$, whose power is model-dependent and qualitatively determines the convergence as the truncation $\dmax$ gets large. The conformal truncation Hamiltonian has sparsity due to various selection rules in the UV CFT. If the UV CFT is a free theory, the most common type of sparsity comes from the number of particles. Take the example of 2D QCD, where we will discuss later that the interaction matrix element can only change the number of particles by 0 or 2. The Hamiltonian is block-wise sparse, and the number of nonzero matrix elements is much smaller than the matrix size. In the large $N_c$ limit, the particle changing process will completely shut down and will restrict to just the $2\rightarrow 2$ block. Global symmetry also results in selection rules. In the 2D QCD case, the theory has a $\mathbb Z_2$-charge conjugation symmetry, which organizes the basis states into even and odd sectors, and the even-odd matrix elements are zero. In more general cases, the basis states are in the irreducible representations of the global symmetry, and the Hamiltonian is sparse due to the selection rule of the Clebsch-Gordan coefficients.

Thus, since it is generic to expect a sparse matrix for the conformal truncation problem, we would expect that conformal truncation is natural for some adiabatic digital algorithms with sparsity, which we will describe in more detail later. 

Finally, we wish to comment on the variational ansatz we need to use for conformal truncation problems. Although it might be related to the exact tasks we are trying to achieve, we do notice a natural variational starting state in the context of conformal truncation. Since we are starting from the UV CFT, and we are using the energy eigenstates from the UV theory, the Hamiltonian is diagonal, and the energy eigenstates will be unit vectors in such a basis. Thus, when we perturb the theory away from the UV fixed point, we could use the starting state as one of the UV eigenstates corresponding to the ground state energy. One could naturally choose a variational ansatz based on such a starting state, and this type of choice will be efficient as long as the perturbation is not large, and we are still in the strongly-coupled regime. In the numerics we show in our examples, we will always use such a variational ansatz, and it is shown to be efficient in our numerical experiments.

\section{Digital quantum simulation}\label{digital}
In this section, we will describe the general theoretical framework of digital quantum simulation in conformal truncation problems. We wish to make this section in general for generic conformal truncation problems, but it also applies to specific models such as 2D QCD. Part of this section is inspired by \cite{kreshchuk2020quantum}.

\subsection{Adiabatic state preparation (ASP)}
We will start with a description of the theory of quantum simulation with a future digital quantum computer. For solving low-lying energy eigenstates and their eigenvalues in a given Hamiltonian, if the Hamiltonian is given by a sum of a relatively simple Hamiltonian $H_0$, and a more complicated Hamiltonian $H_i$, one could consider a dynamical process evolving from an eigenstate of the simpler Hamiltonian $H_0$ to an approximate eigenstate of the more complicated Hamiltonian $H_i$. This is the adiabatic state preparation (ASP) algorithm we will describe here.

The ASP algorithm is based on the adiabatic theorem in quantum mechanics. Say that we have a time-dependent Hamiltonian $H(t)$, and we start from one of its eigenstates, say the non-degenerate ground state $\ket{\Omega (t_0)}$ for simplicity at time $t_0$. During time evolution, we could compute the resulting state $\ket{\psi(t_f)}$ where $\ket{\psi(t_0)}=\ket{\Omega (t_0)}$, on the other hand, we could diagonalize $H(t)$ to get the time-dependent $\ket{\Omega(t)}$. The adiabatic theorem states that if we change the Hamiltonian sufficiently slow and if there is no level crossing, the state $\ket{\psi(t_f)}$ could be very similar to $\ket{\Omega(t_f)}$.

For simplicity, we use the parameter 
\begin{align}
s = \frac{{t - {t_0}}}{{{t_f} - {t_0}}}~,
\end{align}
to parametrize the Hamiltonian, and thus $s\in [0,1]$. The adiabatic theorem says that 
\begin{align}
&\left\| {\left| {\psi \left( t \right)} \right\rangle \left\langle {\psi \left( t \right)} \right| - \left| {\Omega \left( t \right)} \right\rangle \left\langle {\Omega \left( t \right)} \right|} \right\| \le \frac{{\left\| {{H^{(1)}}(0)} \right\|}}{{\Delta t{\Delta ^2}(s)}} + \frac{{\left\| {{H^{(1)}}(s)} \right\|}}{{\Delta t{\Delta ^2}(s)}} \nonumber\\
&+ \frac{1}{{\Delta t}}\int_0^s {\left( {\frac{{\left\| {{H^{(2)}}} \right\|}}{{{\Delta ^2}(s)}} + \frac{{7{{\left\| {{H^{(1)}}} \right\|}^2}}}{{{\Delta ^3}(s)}}} \right)} dx~,
\end{align}
where $\Delta (s)$ is the time-dependent gap of $H(s)$, $\Delta t= t_f-t_i$, and $H^{(i)}$ denotes the $i$-th order derivative of the Hamiltonian over $s$. (See \cite{albash2018adiabatic} for more details.)

Thus, we have a state preparation algorithm for solving the ground state of the Hamiltonian $H(s)$. We could start from $H(0)$ and slowly turn on the Hamiltonian towards $H(1)$ for a sufficiently long period of time, and read the resulting state in the quantum computer as an approximate new ground state. In the quantum simulation of Hamiltonian truncation, it could be used to solve the ground state of the non-conformal theory away from UV, using the conformal data (whose ground state could be directly encoded in a quantum computer). Here, the adiabatic state preparation has a conceptual meaning of renormalization: the time direction of during the adiabatic procedure is the direction of the RG flow. Moreover, in our prescription since we are solving a gapped theory in the middle of the RG flow, we gain a finite amount of $\Delta (s)$ which is independent of system size (it is determined by the nature of the corresponding field theory).

One could easily generalize the above algorithm from the ground state to the low-lying excited states. If one could argue there is no level crossing, it is safe to directly use the statement from the ground state to specific low-lying excited states. An alternative method is that one could compute the first excited state by making it as a ground state for a new Hamiltonian $\tilde{H}(s)=H(s)+\varepsilon_0(s) \ket{\Omega(s)}\bra{\Omega(s)}$ for $s=0$ and $s=1$. Here $\varepsilon_0(s)$ is a large number (larger than the gap of $H(s)$). Then we run the same state preparation algorithm, and we could obtain the first excited state of $H(1)$. One can iterate the above procedure to compute the data for several low-lying excited states. 

Another remark about the above algorithm is the trial to bound the gap rigorously. To make rigorous statements about the required ASP resource (the time we need to spend for the desired error), we have to estimate the upper bound for the gap. One could try analytic methods and classical numerics (from lattice methods, bootstrap methods, or conformal truncation). One could also use the estimated gap obtained from the variational data of the near-term computation for a generic conformal truncation program.

\subsection{Oracle-based algorithms}
Since the ASP algorithm involves a dynamical process $e^{iHt}$, we have to address the following question: given a Hamiltonian at hand, how to construct a set of quantum gates in the quantum computer representing the time evolution $e^{iHt}$?

In the current community of quantum computing, people will mostly focus on simulating quantum many-body systems where the Hamiltonian is given by a sum of local terms. A typical trick one could use, in this case, is the Lie-Trotter formula: one could decompose $e^{i\sum_i H_i}$, where $H_i$ here denotes each local term, as $\prod_i e^{i H_i}$ with extra terms represented by commutators of $H_i$. Complicated higher-order terms in the expansion will involve more sophisticated commutators, but those contributions above the leading term $\prod_i e^{i H_i}$ could be bounded (see the original work \cite{uni} and a very good recent discussion \cite{tro}).

However, for specific problems in conformal truncation, it is hard to naively apply the Lie-Trotter formula since it is purely in the quantum mechanical context where Hamiltonians are non-local. Thus, some specific algorithms might be used.

We again have to take advantage of the structure of conformal truncation problems: we know that the spectrum is very sparse. For sparse Hamiltonians, specific algorithms are designed based on query models and oracles \cite{spar,query} (\emph{black boxes} informing elements of the sparse Hamiltonian to a quantum circuit), showing exponential improvement of efficiency on the precision. We will give a simple introduction to \cite{spar} as the following. 
\begin{theorem}[Efficient simulation for sparse Hamiltonians]\label{simu}
Say that a Hamiltonian is $d$-sparse if there are only $d$ elements in any row or column. Then an $d$-sparse Hamiltonian $H$ in $N=2^n$ dimension could be simulated for time $t$, within error $\varepsilon$ using $\mathcal{O}\left( {\tau \frac{{\log (\tau /\varepsilon )}}{{\log \log (\tau /\varepsilon )}}} \right)$ queries and $O\left( {\tau \frac{{{{\log }^2}(\tau /\varepsilon )}}{{\log \log (\tau /\varepsilon )}}n} \right)$ additional 2-qubit gates, where we define $\tau  \equiv {d^2}{\left\| H \right\|_{\max }}t \ge 1$.
\end{theorem}
We will give a brief explanation of the proof given in \cite{spar} corresponding to the above gate and query complexity. Firstly, we should introduce the query models they use.
\begin{definition}[Discrete-query model]
We start from a bit string $x$. We define $Q_x$ such that
\begin{align}
Q_{x}|j\rangle|b\rangle=e^{-i \pi  b x_{j}}|j\rangle|b\rangle~,~~~\operatorname{for }j \in[N_D] \equiv \{1,2,\cdots,N_D\}\operatorname{and }b \in \{0,1\}~.
\end{align}
We define an algorithm by an arbitrary collection of $x$-independent unitaries and $Q_x$s. We define the discrete-query complexity as the number of $Q_x$ in the full circuit. The discrete-query model is one of the most common quantum query models.
\end{definition}
\begin{definition}[Fractional-query model]
We start from a bit string $x$. We define $Q_x^{\alpha}$ such that
\begin{align}
Q_{x}^{\alpha}|j\rangle|b\rangle=e^{-i \pi \alpha b x_{j}}|j\rangle|b\rangle~,~~~\operatorname{for }j \in[N_F] \equiv \{1,2,\cdots,N_F\}\operatorname{and }b \in \{0,1\}~.
\end{align}
We define an algorithm in the fractional query model, as a collection of unitaries 
\begin{align}
U_{m} Q_{x}^{\alpha_{m}} U_{m-1} \cdots U_{1} Q_{x}^{\alpha_{1}} U_{0}~.
\end{align}
Here $U$s are arbitrary unitaries and $\alpha_{i} \in(0,1]$. We see that the fractional-query complexity of the algorithm is $\sum_{i=1}^{m} \alpha_{i}$, and $m$ is the number of gates appearing in the algorithm.
\end{definition}
\begin{definition}[Continuous-query model]
We also start from a bit string $x$. We define $H_x$ such that
\begin{align}
H_{x}|j\rangle|b\rangle=\pi b x_{j}|j\rangle|b\rangle~,~~~\operatorname{for }j \in[N_C] \operatorname{and }b \in \{0,1\}~.
\end{align}
An algorithm in the continuous-query model is given by an $x$-independent Hamiltonian $H_D(t)$ driving in the time interval $t \in[0, T]$. The algorithm implements the unitary $U(T)$ by the time evolution given by $H_D(t)$
\begin{align}
i \frac{{d}}{{d} t} U(t)=\left(H_{x}+H_{D}(t)\right) U(t)~.
\end{align}
Thus the continuous-query complexity is given by the total time $T$.
\end{definition}
Based on the above query models, the proof is given by the following two steps: continuous-query simulation and Hamiltonian simulation reduction. 

For continuous-query simulation, the starting point will be the following theorem,
\begin{theorem}
Consider an algorithm in the fractional-query model. If the fractional-query complexity is 1 (or less than 1), then we could implement it in the discrete-query model with queries $\mathcal{O}\left(\frac{\log (1 / \varepsilon)}{\log \log (1 / \varepsilon)}\right)$, with error at most $\varepsilon$.
\end{theorem}
This theorem shows the connection between the fractional-query model and the discrete-query model. Furthermore, we could prove the equivalence between the continuous-query model and the fractional-query model.
\begin{theorem}
For any $\varepsilon>0$, if we have an algorithm in the fractional-query model with fractional-query complexity $T$, we could implement it in the continuous-query model with continuous-query complexity $T$, with error at most $\varepsilon$. Furthermore, if we have an algorithm in the continuous-query model with continuous-query complexity $T$, we could implement it in the fractional-query model with continuous-query complexity $T$, with error at most $\varepsilon$ and $m=\mathcal{O}\left(\bar{h} T^{2} / \varepsilon\right)$ fractional-query gates, where
\begin{align}
\bar{h}\equiv \frac{1}{T} \int_{0}^{T}\left\|H_{D}(t)\right\|dt ~,
\end{align}
is the average norm of the corresponding driven Hamiltonian. 
\end{theorem}
Now, we arrive at the main theorem of the continuous-query simulation:
\begin{theorem}[Continuous-query simulation]
If we have an algorithm with continuous(or fractional)-query complexity $T$, and we assume $T\ge 1$, then we could simulate it using 
\begin{align}
\mathcal{O}\left( {T\frac{{\log (T/\varepsilon )}}{{\log \log (T/\varepsilon )}}} \right)
\end{align}
queries with error $\varepsilon$. Furthermore, in the case of the continuous-query model, if we assume the time evolution of the Hamiltonian $H_D(t)$ could be implemented using $g$ 2-qubit gates with error $\varepsilon/T$ between any two times, then the estimate of total number gates is 
\begin{align}
\mathcal{O}\left( {T\frac{{\log (T/\varepsilon )}}{{\log \log (T/\varepsilon )}}[g + \log (\bar hT/\varepsilon )]} \right)~,
\end{align}
where we define 
\begin{align}
\bar{h}\equiv \frac{1}{T} \int_{0}^{T}\left\|H_{D}(t)\right\|dt~.
\end{align}
\end{theorem}
Then we could move to Hamiltonian simulation reduction. We firstly reduce a given $d$-sparse Hamiltonian to a sum of 1-sparse Hamiltonians. Then, we could use the Lie-Trotter formula and understand each term in the product as oracles. After several simplifications, one could arrive at the efficient simulation result, Theorem \ref{simu}. 

Now one might ask, what is the proper treatment when we consider a time-dependent situation? This is particularly important for us since we need to combine adiabatic state preparation and sparse Hamiltonian simulation together. One could generalize the above algorithm based on the fractional query model naively to the time-dependent case \cite{spar}. However, one could use the following updated algorithm designed specifically for the time-dependent Hamiltonian evolution \cite{berry2019time}.
\begin{theorem}[Efficient simulation for time-dependent sparse Hamiltonians]\label{simutime}
Say that a time-dependent Hamiltonian $H(\tau)$ is at most $d$-sparse during a time interval $[0,t]$ in $2^n$ dimensions. Then it could be simulated for time $t$, within error $\varepsilon$ using 
\begin{align}
O\left( {d{{\left\| H \right\|}_{\max ,1}}\frac{{\log \left( {d{{\left\| H \right\|}_{\max ,1}}/\varepsilon } \right)}}{{\log \log \left( {d{{\left\| H \right\|}_{\max ,1}}/\varepsilon } \right)}}} \right)
\end{align}
queries and 
\begin{align}
\tilde{O}\left( {d{{\left\| H \right\|}_{\max ,1}}n} \right)
\end{align}
additional 2-qubit gates, where we define 
\begin{align}
{\left\| H \right\|_{\max ,1}} \equiv \int_0^t {{d}} \tau {\left\| {H(\tau )} \right\|_{\max ,1}}~,
\end{align}
as the integral of the maximal eigenvalue during the time evolution.
\end{theorem}

The notation $\tilde{O}$ means that we are ignoring the logarithmic dependence in this formula. In this result, there is a further improvement on the dependence of sparsity from $d^2$ to $d$. This result (Theorem 10 in \cite{berry2019time}) is also based on the idea of using the Dyson series and Taylor expansion, which is partially from \cite{updatespar}.

The detailed implementation of the algorithms is given in \cite{spar,berry2019time} more precisely\footnote{See some similar oracle-based algorithms, for instance, \cite{wan2020fast}.}. Since the algorithms work for general Hamiltonians with sparsity, we could make a direct analysis of the case we are looking at, the conformal truncation.

\subsection{Analysis in conformal truncation}
We present a brief analysis of applying those algorithms to conformal truncation. Combining the spirit of adiabatic state preparation and sparse Hamiltonian simulation, for conformal truncation problems,
\begin{align}
H = {H^{{\rm{CFT}}}} + gV~,
\end{align}
we could take 
\begin{align}
H(s) = (1 - s){H^{{\rm{CFT}}}} + gsV~,
\end{align}
where $s=\tau/t\in [0,1]$. Thus we have
\begin{align}
\int_0^t d \tau {\left\| {H(\tau )} \right\|_{\max ,1}} = t\int_0^1 d s{\left\| {H(s)} \right\|_{\max ,1}} \equiv t\left[\kern-0.15em\left[ H 
 \right]\kern-0.15em\right]~.
\end{align}
The size of the norm might be harmful to our simulation. However, we could always rescale the Hamiltonian to make the norm more suitable for our calculations. Moreover, sparsity is particularly important for our simulation. For 2D QCD, the sparsity scales as half of the truncation $\Delta_{\max}$. In more general models, the sparsity might be even smaller, scaling as $\log \Delta_{\max}$ or even less. Thus we gain a remarkable advantage based on sparsity. Another feature of this algorithm is an exponential improvement in precision, which is particularly impressive as a feature of oracle-based algorithms. 

\section{Target: 2D QCD}\label{QCD}
We briefly review the 2D QCD in this section. Historically 2D QCD at large $N_c$, also known as the `t Hooft model, is solved by `t Hooft in 
\cite{t1993two}, where he showed confinement analytically. For a thorough review, see also 
\cite{callan1993two}. In the context of Hamiltonian truncation, the model is studied in 
\cite{Pauli:1985ps,Hornbostel:1988fb}
using the discretized light-front quantization (DLCQ) method. In this paper, we will use the {\it Lightcone Conformal Truncation} (LCT) framework. In 
\cite{Katz:2014uoa}, people study 2D QCD with massless quarks at any $N_c$ in this framework. The result agrees with DLCQ with a much smaller Hilbert space.

In this section, we introduce the model 2D QCD and setup the conformal truncation for the model. The section follows the work \cite{pedagogical}\footnote{We thank Nikhil Anand, Liam Fitzpatrick, Emanuel Katz, Zuhair Khandker, Matthew Walters for letting us use the preliminary results.}.

\subsection{Integrate out the gauge field}

We begin with the Yang-Mills Lagrangian in $(1+1)$ dimension
\begin{equation}
  \mathcal{L}_{\rm{YM}} = -\frac{1}{2} \Tr F_{\mu\nu} F^{\mu\nu} - A_\mu J^\mu~,
\end{equation}
where the gauge field is in the adjoint representation of $\text{SU}(N_c)$ group. In lightcone quantization, we choose the lightcone coordinate $x^{\pm} \equiv (t\pm x)/ \sqrt2$, where $x^+$ is the lightcone \emph{time} and $x^-$ is the \emph{space}. The Lagrangian can be written as
\begin{equation}
  \mathcal{L}_{\rm YM } =  \Tr (\d_- A_+)^2 - A_+ J_-~,
  \label{eq:gaugethrylag}
\end{equation}
where we choose the lightcone gauge $A_- = 0$. Notice that the equation of motion of the remaining component of the gauge field, $A_+$ is non-dynamical since there is no $x^+$ derivative, 
\begin{equation}
A_+^a = -\frac{1}{\d_-^2} J_-^a ~.
\end{equation}
We can completely integrate out the field $A_+$ by substituting its equation of motion and get rid of the gauge field
\begin{equation}
  \mathcal{L}_{\rm YM } = \frac{1}{2} \cdot J^a \frac{1}{\d^2} J^a ~.
  \label{eq:YMlag}
\end{equation}
In 2D QCD, the full Lagrangian is the gauge field coupled to the quarks, which are in the fundamental representations
\begin{equation}
  \mathcal{L}_{\rm QCD} = \bar\Psi (i \slashed{D}-m) \Psi - \frac{1}{2} \Tr F_{\mu\nu}F^{\mu\nu}~.
\end{equation}
For simplicity we consider the massless case $m=0$. The current in (\ref{eq:gaugethrylag}) and (\ref{eq:YMlag}) is thus
\begin{equation}
  J^a = g \bar{\Psi} \gamma_- T^a \Psi ~.
\end{equation}
We take the following convention for the 2D fermion field
\begin{align}
&\Psi = \left( \begin{array}{c} \chi \\ \psi \end{array} \right)~, \quad \bar{\Psi} = \left( \psi^\dagger,\chi^\dagger \right)~,\nonumber\\
&\gamma^+ = \gamma_- =\begin{pmatrix}
    0 & 0 \\
    \sqrt{2} & 0
  \end{pmatrix}~, \quad \gamma^- = \gamma_+ = \begin{pmatrix}
    0 & \sqrt{2}\\
    0 & 0
  \end{pmatrix}~, \quad \gamma^0 = \begin{pmatrix}
    0 & 1 \\
    1 & 0
  \end{pmatrix}~.
\end{align}
With a little bookkeeping, we fill find that the $\chi$ component has a trivial equation of motion 
\begin{equation}
\d_- \chi = 0~,
\end{equation}
and can be integrated out. The remaining Lagrangian is thus
\begin{equation}
  \mathcal{L} \equiv \mathcal{L}^{\rm (CFT)} + \delta \mathcal{L}
  = i \sqrt{2} \psi \d_+ \psi 
  + g^2 \psi^\dagger T^a \psi \frac{1}{\d_-^2} \psi^\dagger T^a \psi ~,
  \label{eq:2dQCDLag}
\end{equation}
where $\mathcal{L}^{\rm (CFT)} \equiv i \sqrt{2} \psi \d_+ \psi$ describes the CFT of a chiral fermion, and $\delta \mathcal{L}$ describes the gauge interaction. In the last equation we suppress the contracted color indices
\begin{equation}
\psi_i^\dagger (T^a)_{ij} \psi_j~.
\end{equation}

The Hamiltonian is the generator of the lightcone \emph{time} translation, i.e., the $P_+$ operator. The mass shell relation is $2 p_+ p_- = \mu^2$ where $\mu$ is the mass or center-of-mass energy. Using Lorentz invariance, we can boost any state to $p_- = 1$ and states with different $p_-$ never mix due to momentum conservation, so we only need to consider the $p_- = 1$ sector of the Hilbert space.
We conveniently multiply the Hamiltonian by a $(2p)$ factor, so that the Hamiltonian measures the mass.
We obtain the Hamiltonian through a Legendre transformation,
\begin{equation}
H \equiv 2 p P_+ = (2p) (-g^2) \int dx^-
: \psi^\dagger T^a \psi \frac{1}{\d_-^2} \psi^\dagger T^a \psi : 
\, (x^-)~,
\label{eq:2dQCDHam} 
\end{equation}
where $ H^{\rm (CFT)} = 0$ due to the fact that the equation of motion of free fermion $\d_+ \psi = 0$.
From now on we write $\d \equiv \d_-$ (and hence $p \equiv p_-$) because $\d_+$ kills the operator (and hence $p_+ = 0$ for the basis states).

\subsection{The UV CFT and the conformal basis}

As we discussed in Section \ref{sec:introToConformalTruncation}, in conformal truncation the basis is constructed from the local operators in the UV CFT,
\begin{equation}
\tag{\ref{eq:basis}}
| \CO, p \> \equiv \int e^{i p x} dx \, \CO(x) | 0 \> ~.
\end{equation}
Because the operator is independent of $x^+$ from now on, we write $x \equiv x^-$. Our goal is to find a complete orthonormal basis up to $\dmax$ for the specific UV CFT.
For 2D QCD (\ref{eq:2dQCDHam}) the UV CFT is a free complex fermion field $\psi$ valued in the fundamental representation of $\text{SU}(N_c)$
\begin{equation}
  \begin{aligned}
    \psi_j(x) = \int \frac{dp}{\sqrt{8\pi^2}} \left({e^{-ipx} b_{j} + e^{i p x}a_{j} ^\dagger}\right) ~,\\
    \psi_j^\dagger(x) =  \int \frac{dp}{\sqrt{8\pi^2}}\left( {e^{ipx} b^\dagger_{j} + e^{-i p x}a_{j}} \right) ~,\label{eq:complexfermionmodefunctions}
  \end{aligned}
\end{equation}
where the creation and annihilation operators satisfy 
\begin{equation}
  \begin{aligned}
  \{ a_i(q), a^\dagger_j(p) \} &= \delta_{ij} (2\pi)\delta(q - p)~, \\
    \{ b_i(q), b^\dagger_j(p) \} &= \delta_{ij} (2\pi)\delta(q - p)~.
  \end{aligned}
\end{equation} 
It can be proven that a complete set of gauge-invariant local operators constructed from $\psi$ and $\psi^\dagger$ have the form 
\begin{equation}
\CO(x) =  \sum_{\sum k_i = \Delta - n} 
c_{k_1, k_2, \cdots, k_{2n}} 
\pr{ \d^{k_1} \psi_{i_1}^\dagger \d^{k_2} \psi_{i_1} }
\cdots
\pr{ \d^{k_{2n-1}} \psi_{i_n}^\dagger \d^{k_{2n}} \psi_{i_n} }~,
\label{eq:UVOperator}
\end{equation}
where $\Delta$ is the scaling dimension of the operator, and the operator creates a state of $n$ fermions and $n$ anti-fermions. The coefficients have to satisfy the condition that the sum is a primary operator, (\ref{eq:primCondition}).
In practice, we construct the operator recursively using a result obtained by Penedones in 
\cite{penedones2011writing}
\footnote{See also earlier work by Mikhailov
\cite{Mikhailov:2002bp}.
}. The result states that for primary operators $A$ and $B$, one can build a tower of \emph{double trace} operators that are also primary
\begin{equation}
A \overleftrightarrow{\d}^\ell B \equiv 
\sum_{k=0}^\ell c_\ell^k (\Delta_A, \Delta_B) 
~ \d^{k} A \, \d^{\ell - k} B~,
\label{eq:doubleTraceConstruction}
\end{equation}
where the coefficients are
\begin{equation}
c^k_\ell(\Delta_A,\Delta_B) = \frac{(-1)^k \Gamma(2\Delta_A+\ell) \Gamma(2\Delta_B+\ell)}{k! (\ell-k)! \Gamma(2\Delta_A+k) \Gamma(2\Delta_B + \ell - k)} ~.
\label{eq:doubleTraceCoeff}
\end{equation}
If both $A$ and $B$ are generalized free fields, then there will be exactly one primary operator for each non-negative integer $\ell$ in \ref{eq:doubleTraceConstruction}. In our construction, we can generate an over-complete basis by constructing the double trace operators for all pairs of lower-dimensional primary operators, then perform Gram-Schmidt orthogonalization. 

The inner product between basis states is defined through the spacial two-point function of the UV CFT
\begin{align}
\langle \CO, p | \CO^\prime, p^\prime \rangle & \equiv 
\int dx \, dy \, e^{i(px - p^\prime y)} \langle \CO(x) \CO^\prime(y) \rangle  
\tag{\ref{eq:InnerProduct} }
\\
& = 
2p(2\pi) \delta(p-p^\prime) \, G_{\CO\CO^\prime}~, \nonumber
\end{align}
where $\CO$ and $\CO^\prime$ are un-normalized primary operators. 
We choose our normalization to be such that $G_{\CO\CO^\prime} = \delta_{\CO\CO^\prime}$.
The spacial two-point function of the UV CFT takes the general form
\begin{equation}
\langle \CO(x) \CO^\prime(y) \rangle = 
\frac{\mathfrak{g}_{\CO\CO^\prime}}{(x-y)^{2\Delta}} ~,
\end{equation} 
where $\mathfrak{g}_{\CO\CO^\prime}$ is known as the Zamolodchikov metric. Thus choosing an orthonormal basis for the Hilbert space is equivalent to diagonalizing the CFT two-point function, up to an overall $\Delta$-dependent factor
\begin{equation}
G_{\CO\CO^\prime} = 
 \frac{1}{2p}
\int dx \, \frac{e^{ipx}}{x^{2\Delta}} 
\mathfrak{g}_{\CO\CO^\prime}
= \frac{\pi p^{2\Delta-2}}{\Gamma(2\Delta)}
\mathfrak{g}_{\CO\CO^\prime}~.
\end{equation}
The $\mathfrak{g}_{\CO\CO^\prime}$ can be computed using the free fermion two-point function as a building block
\begin{equation}
\corr{\d^{k} \psi_i^\dagger(x) \d^{k'} \psi_j(y)} =\frac{\Gamma(k+k'+1)}{4\pi(x-y)^{k+k'+1}}  \delta_{ij}~. \label{eq:gen2ptfunctionQCD}
\end{equation}
Since the fermions are free in UV, the two-point function of composite operators as products of $\psi$'s and $\psi^\dagger$'s factorize. We can, therefore, expand the operators as (\ref{eq:UVOperator}), and for each term, sum up all possible {\it Wick contractions}, and obtain $\mathfrak{g}_{\CO\CO^\prime}$.

\subsection{Compute the matrix element}

We compute the matrix elements of the Hamiltonian with respect to the basis states using the three-point function of the UV CFT
\begin{align}
\langle \CO,p | H |  \CO^\prime, p^\prime \rangle &\equiv 
\int dx \, dy \, dz\, e^{i(px - p^\prime z)} 
\langle \CO(x) \CO_R(y) \CO^\prime(z) \rangle 
\tag{\ref{eq:MatrixElements}}\\
& = 
2p (2\pi) \delta(p-p^\prime)  \CM_{\CO\CO^\prime}~, \nonumber
\end{align}
for the 2D QCD Hamiltonian (\ref{eq:2dQCDHam}), the deformation is $\CO_R(y) \equiv \psi^\dagger T^a \psi \frac{1}{\d^2} \psi^\dagger T^a \psi$. 

For the three-point function part, we use Wick contraction again, taking the building blocks (\ref{eq:gen2ptfunctionQCD}). There are a few new objects. First, we use the identity
\begin{equation}
(T^a)_{ij} (T^a)_{mn} = \half\pr{\delta_{in}\delta_{j m} - \frac{1}{N_c} \delta_{ij}\delta_{mn}}~. \label{eq:sunidentity}
\end{equation}
After we contract out the $\psi$'s and $\psi^\dagger$'s, there should be no free color indices left, and the tensor contractions will give $N_c$ factors. Second, we need an appropriate definition for the non-local potential $\frac{1}{\d}$. The most transparent definition of $\frac{1}{\d}$ is $\frac{1}{p}$ in the momentum space, thus we Fourier transform the spacial factors like $(x-y)^{-k}$ 
and proceed with (\ref{eq:MatrixElements}) as a momentum space integral. Schematically the momentum integrand corresponds to the following Feynmann diagram,
\begin{align}\label{eq:gauge-interaction-fock}
(\psi^\dagger_i T_{ij}^A \psi_j) \frac{1}{\d^2} (\psi^\dagger_k T_{kl}^A \psi_l)
~\sim~ 
\raisebox{-.40in}{\includegraphics[width=1.3in]{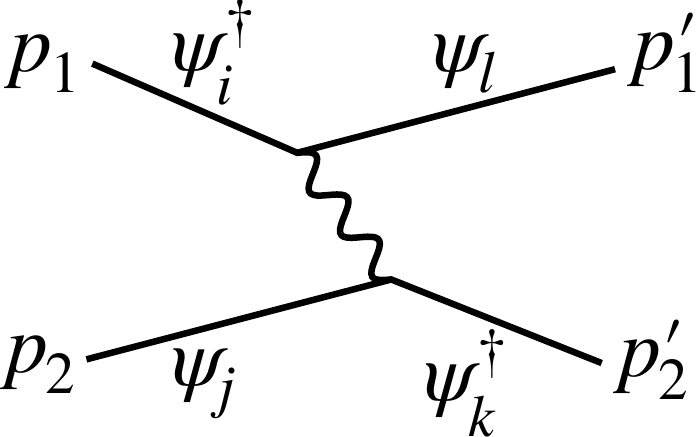} }~
\sim~& \frac{1}{(p_1-p_1^\prime)^2}~,
\end{align}
where only the momentum factor coming from the $\frac{1}{\d^2}$ is shown. 
Finally, the momentum integral in (\ref{eq:gauge-interaction-fock}) has an IR divergence when $p_1 \rightarrow p_1^\prime$. The divergence reflects an ambiguity of normal ordering (\ref{eq:2dQCDHam}). The normal order gives a self-energy shift term 
\begin{align}\label{eq:gauge-self-energy}
\raisebox{-.2in}{\includegraphics[width=1.2in]{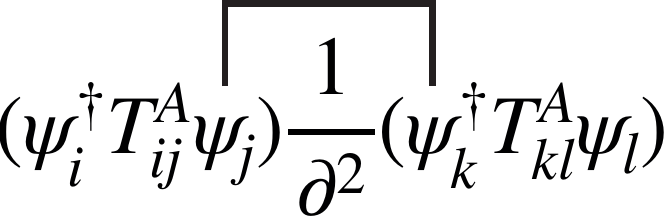} } 
~\sim~ 
\raisebox{-.3in}{\includegraphics[width=1.4in]{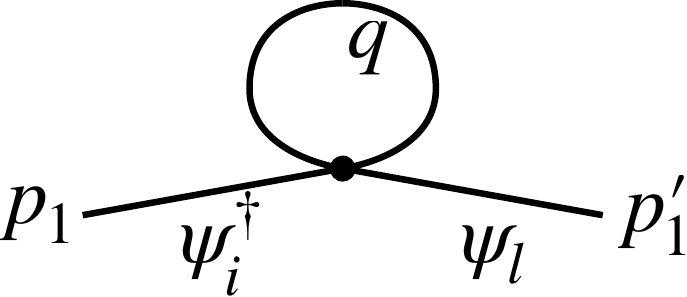} } 
\sim \int \frac{dq}{(p_1-q)^2} (2\pi)\delta(p_1-p_1^\prime) ~,
\end{align}
which has the same IR divergence. The divergence cancels, in a way that chiral symmetry is manifest, and the lowest eigenstate, pion, is exactly massless. For the technical details of the computation of the matrix elements, see Appendix \ref{sec:QCDappendix}.

\subsection{Large \texorpdfstring{$N_c$}{Nc} limit}

The problem is greatly simplified in the limit $N_c \rightarrow \infty$ keeping the `t Hooft coupling $\lambda \equiv g^2 N_c$ finite. By simply counting the $N_c$ factors from contracting the color indices in the normalized matrix elements 
\begin{align}
\CM_{2n \rightarrow 2n}&\equiv 
\frac{\<\CO_{2n}|H|\CO_{2n}^\prime\>}{
\sqrt{\<\CO_{2n}|\CO_{2n}\> } 
\sqrt{\<\CO_{2n}^\prime|\CO_{2n}^\prime\>} }
\sim g^2 N_c~, \nonumber \\
\CM_{2n \rightarrow 2n+2}&\equiv 
\frac{\<\CO_{2n}|H|\CO_{2n+2}^\prime\>}{
\sqrt{\<\CO_{2n}|\CO_{2n}\> } 
\sqrt{\<\CO_{2n+2}^\prime|\CO_{2n+2}^\prime\>} }
\sim g^2 \sqrt{N_c}~,
\end{align}
we see that the particle-changing matrix elements are suppressed, and we can restrict the Hilbert space to the two-quark sector.  

The two-quark basis states are, following 
(\ref{eq:doubleTraceConstruction}) and
(\ref{eq:doubleTraceCoeff})
\begin{equation}
| \CO_\ell, P \> =
\frac{1}{A_\ell} 
\int e^{i p x} dx \ 
 \psi^\dagger \overleftrightarrow{\d}^\ell \psi (x) | 0 \>~,
 \qquad
 \ell = 0, 1, 2 \cdots~,
\end{equation}
where 
\begin{equation}
A_\ell^2 = \frac{p^{2\ell}N_c}{(2\ell+1) 16\pi }~,
\end{equation} is the normalization factor and
\begin{equation}
\psi^\dagger \overleftrightarrow{\d}^\ell \psi (x)
\equiv \sum_{k=0}^\ell \frac{(-1)^k\Gamma(\ell+1)^2}{k!(\ell-k)!\Gamma(k+1)\Gamma(\ell-k+1)} \,
\d^k \psi^\dagger_i \d^{\ell-k}\psi (x)~,
\end{equation} 
is the primary operator. Schematically, the basis states describe a free quark and a free anti-quark moving with different relative momenta. The letter $\ell$ labels the wave function of relative momenta, where larger $\ell$ describes the further separation between the pair.
Once we add the gauge interaction, the free quarks confine, and the physical spectrum contains mesons formed by a pair of quark and anti-quark. Since pair production and annihilation are turned off in the large $N_c$ limit, each meson contains exactly one quark and one anti-quark. Since there is no free quark in the physical spectrum, in each meson eigenstate, the separation between the pair has to be bounded. Hence the low energy eigenstates should have low overlap with basis states of high $\ell$, and the result should rapidly converge as we increase $\Delta_{\rm max}$.

The matrix elements are 
\begin{equation}
\CM_{\ell\ell^\prime} \equiv 
\frac{\<\CO_{\ell}|H|\CO_{\ell^\prime}\>}{
\sqrt{\<\CO_{\ell}|\CO_{\ell}\> } 
\sqrt{\<\CO_{\ell^\prime}|\CO_{\ell^\prime}\>} }~,
\end{equation}
where the factor $(2p)$ can be organized as an integral
\begin{equation}
\begin{aligned}
&\CM_{\ell\ell^\prime} 
= 
 \frac{g^2 N_c}{\pi}
 \sqrt{2\ell+1}\sqrt{2\ell^\prime+1}
   ~,\\
&\text{PV}
\int_{0}^{1} dx_1 dx_2
   P_{\ell_1}\left(1-2
   x_1\right) 
   \frac{
    P_{\ell_2}\left(1-2 x_1\right)
    - P_{\ell_2}\left(1-2 x_2\right)
   }{\left(x_1-x_2\right){}^2}~,
\end{aligned}
\label{eq:largeNMatrixElementsIntegral}
\end{equation}
where 
\begin{equation}
P_{\ell}(w) \equiv \frac{1}{2^\ell \ell!} \frac{d^\ell}{dw^\ell} (w^2-1)^\ell~,
\end{equation} 
is the Legendre polynomial
and PV stands for the principal value prescription.

For example, the matrix with a small truncation $\dmax = 4$ is shown as
\begin{equation}
H = 
\frac{g^2 N_c}{\pi}\pr{\begin{matrix}
0 & 0 & 0 & 0 \\
0 & 6 & 0 & \sqrt{\frac{7}{3}} \\
0 & 0 & 15 & 0 \\
0 & \sqrt{\frac{7}{3}} & 0 & \frac{77}{3}
\end{matrix}}~,
\end{equation}
where the states from the first to last are $\ell = 0,1,2,3$, respectively. We see one exactly massless state, which is the massless pion due to chiral symmetry. The matrix elements are zero unless $\ell$ and $\ell^\prime$ are either both odd or both even. This is because the model respects charge conjugation symmetry. The $\ell$ even states have odd parity, and $\ell$ odd states have even parity under charge conjugation.

\section{Variational quantum simulation}\label{variational}
\subsection{Variational quantum eigensolver (VQE)}
We generically describe the theory of variational quantum eigensolver. The variational quantum eigensolver (VQE) \cite{peruzzo2014variational,farhi2014quantum,mcclean2016theory} is based on the variational principle of quantum mechanics. Say that we have a Hamiltonian $H$ with $N$ qubits. Namely, the dimension of the Hilbert space $\mathcal{H}$ is $\text{dim}\mathcal{H}=2^N$. Now we wish to find the low-lying spectrum of the Hamiltonian $H$ and their corresponding energy eigenstates. 

Now, if we have a quantum computer, we could consider prepare a set of states $V = {\{ \psi (\vec \theta )\} _{\vec \theta }}$, where $\vec{\theta}$ is a $q$-dimensional vector representing parameters, which we call the space of ansatz. We hope to choose $V$ to be significantly smaller than the whole Hilbert space $\mathcal{H}$. 

The quantum computer is assumed to be able to prepare the states in $V$ efficiently. For instance, one could assume that the space $V$ could be given by states in the form
\begin{align}
\left| \psi (\vec{\theta})  \right\rangle  = {U_q}({\theta _q}) \ldots {U_2}({\theta _2}){U_1}({\theta _1})\left| {{\psi _{{\rm{simple}}}}} \right\rangle ~,
\end{align}
where $U_i$s are simple unitary operations, for instance, changing the coupling of local spins in the Hamiltonian. $\psi_\text{simple}$ are states that are very easy to prepare, for instance, product states in spin systems. This type of $V$ is pretty common in experiments of quantum simulation for current technologies, such as trapped-ion systems \cite{Kokail:2018eiw}. We could also include further entanglement between nearby qubits in the variational ansatz.

For a given $\vec{\theta}$, one could measure the expectation value of some operators in the quantum devices. We could compute the cost function $F(\vec \theta ) $ as a function of the expectation values of the corresponding operators. For instance, one could choose $F(\vec \theta ) = {\left\langle H \right\rangle _{\vec \theta }}$, the expectation value of the Hamiltonian itself, or $F(\vec \theta ) = {\left\langle {{H^2}} \right\rangle _{\vec \theta }} - \left\langle H \right\rangle _{\vec \theta }^2$, the variance of the Hamiltonian. One could use some numerical algorithms to search for an optimal $\vec{\theta}$ that could minimize the cost function based on the result of measurements. For numerical optimization in higher dimensions, one could consider, for instance, the gradient descent method,
\begin{align}
\vec \theta  + d\vec \theta  = \vec \theta  - \gamma \vec \nabla F~,
\end{align}
for sufficiently small $\gamma$. All those computations could be done in a classical device. Thus usually, VQE is performed in a hybrid quantum-classical way, which is more practical for near term devices. 

The efficiency of VQE is often hard to estimate theoretically since its efficiency depends significantly on the structure of the system itself, and the space of ansatz $V$ we choose. Thus, classical simulators for quantum devices are useful to probe the efficiency and scalability of the quantum simulation tasks.

When performing VQE, the simplest task one could consider is to determine the ground state. If the states we are interested in are not the ground state, but the second low-lying excited state, one could consider modifying the Hamiltonian by adding $\alpha \ket{\psi_\text{ground}} \bra{\psi_\text{ground}}$, where $\alpha$ is sufficiently large (at least larger than the estimated gap of the Hamiltonian). Then the corresponding excited state becomes the ground state for the new Hamiltonian. Using this method, one can, in principle, solve the spectrum of the Hamiltonian by performing a sufficiently large number of VQEs \cite{spec}.

\subsection{Encoding}
Here we briefly discuss the encoding schemes we will use. Let us assume that we have a tower of states $\ket{s}$ from 0 to $d-1$. The state should, in principle, be represented and operated in the qubit system. There are two most common ways of doing the encoding: the direct mapping and the compact mapping (see introduce for instance, \cite{kreshchuk2020quantum,somma2005quantum}). 

The direct mapping maps the state $\ket{s}$ as
\begin{align}
|s\rangle  =  \otimes _{j = 0}^{s - 1}|0{\rangle _j}|1{\rangle _s} \otimes _{j = s + 1}^{d - 1}|0{\rangle _j}~.
\end{align}
In the Hamiltonian, the creation operator maps to 
\begin{align}
{a^\dag } = \sum\limits_{s = 0}^{d - 2} {\sqrt {s + 1} } |0\rangle {\left\langle 1 \right|_s} \otimes \left| 1 \right\rangle {\left\langle 0 \right|_{s + 1}}~.
\end{align}
In this approach, we have to use $d$ qubits, with total Hilbert space dimension $2^d$. It is very expensive in system size, but the advantage is that the creation operator could be mapped to a sum of 2-local terms. 

There is another way, which is called compact mapping. In this case, we transform the number $s$ to its binary representation
\begin{align}
s = {b_{K - 1}}{2^{K - 1}} + {b_{K - 2}}{2^{K - 2}} +  \ldots {b_0}{2^0}~,
\end{align}
and we could write down the mapping for the state as
\begin{align}
|s\rangle  = \left| {{b_{K - 1}}} \right\rangle \left| {{b_{K - 2}}} \right\rangle  \ldots \left| {{b_0}} \right\rangle ~,
\end{align}
where $K = \left\lfloor {\log d} \right\rfloor $. In this framework, one could write down the representation of the creation operator
\begin{align}
a^{\dagger}=\sum_{s=0}^{d-2} \sqrt{s+1}|s+1\rangle\langle s|~.
\end{align}
In this method, we could use $K = \left\lfloor {\log d} \right\rfloor $ qubits, but we don't have the locality in the creation operator. 

Typically, the choice of encoding schemes depends on situations of tasks and quantum devices. In this work, we will use the compact encoding scheme for our illustrative calculations. 
\subsection{Trial numerics}
\begin{figure}[H]
  \centering
  \includegraphics[width=0.6\textwidth]{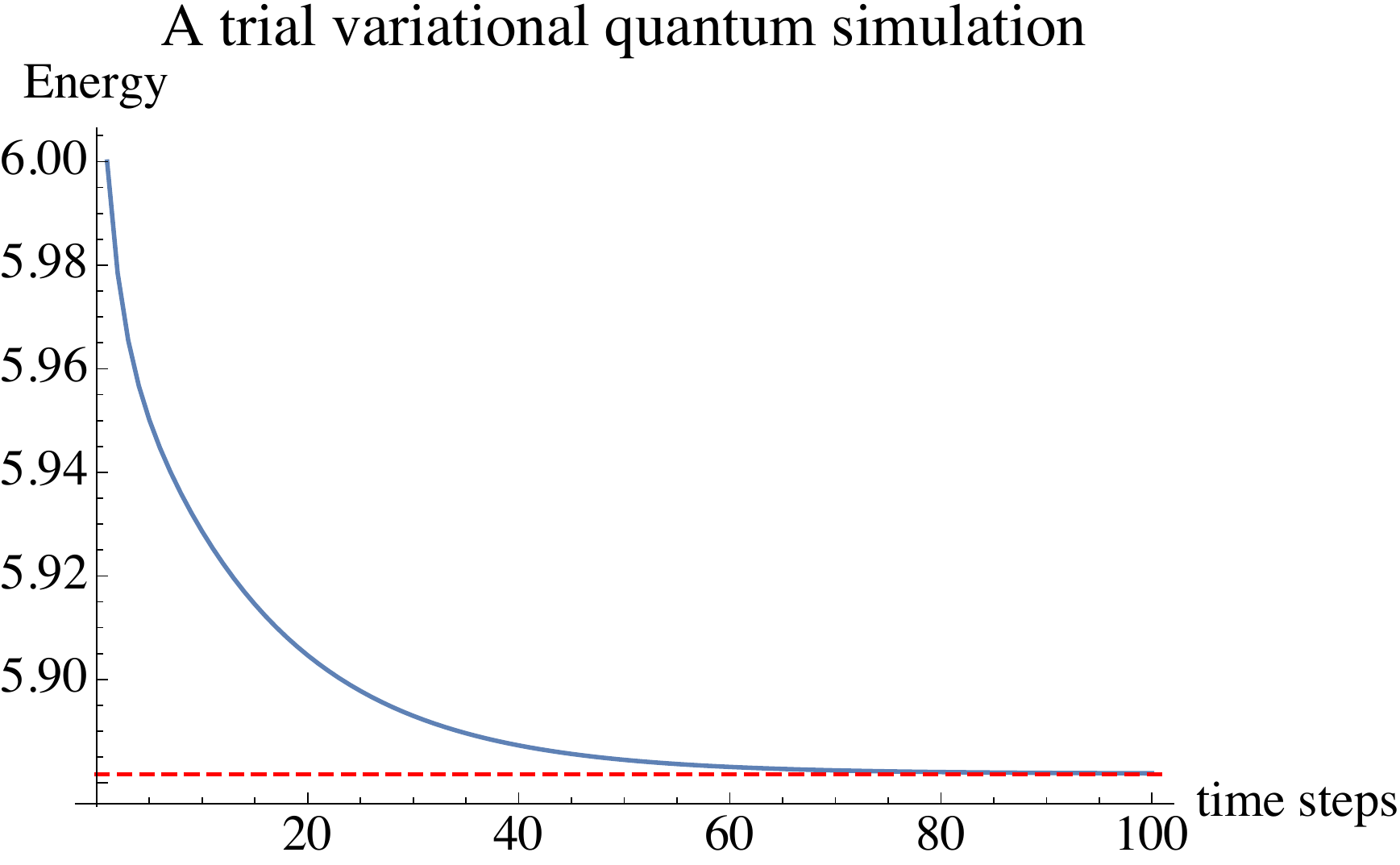}
  \caption{\label{fig1} A trial for variational quantum simulation. We run a simple VQE computation in a classical setup without any noise. We use $\gamma=0.00006$ for gradient descent method, and we run the simulation for 100 steps. Blue: VQE result. Red dashed: exact result.}
\end{figure}

A starting point of the numerical simulation is to run trial numerics about the variational calculation of the first excited state in 2D QCD in the large $N_c$ limit. In Figure \ref{fig1}, we present a simple numerics by truncating the matrix as $\Delta_{\max}=9$. Because there is always a massless pion, we drop the first column and the first row in the matrix. So effectively, it has the dimension $8\times 8$. Now, the first excited state (a meson state) effectively becomes a ground state after we drop out the massless pion. It has an energy between 5.8 and 5.9 after we drop the overall constant, $g^2N_c/\pi$. In the $\Delta_{\max}=9$ truncation, the exact answer is 5.8817. Now, we solve this energy by variational quantum simulation. We start from the corresponding eigenstate in UV CFT $\ket{\psi_{\text{CFT}}}$, which has energy 6. Then we evolve the state by the variational ansatz
\begin{align}
|\psi (\vec \theta )\rangle  = \exp (i{\theta _1}{Y_1})\exp (i{\theta _2}{Y_2})\exp (i{\theta _3}{Y_3})\left| {{\psi _{{\rm{CFT}}}}} \right\rangle ~,
\end{align}
where we encode the system into a 3-qubit quantum setup using compact encoding and $Y_i$s are Pauli $Y$s. We use the cost function
\begin{align}
F(\vec \theta ) = {\left\langle {{H^2}} \right\rangle _{\vec \theta }} - \langle H\rangle _{\vec \theta }^2~,
\end{align}
and after 100 steps, we obtain a good convergence to the exact result. Note that we use the simplest variational ansatz does not contain any entanglement. For more complicated problems, some other variational ansatz might be used. 

\subsection{\texttt{IBM Quantum Experience}}
To justify our method, we perform some numerical experiments in \texttt{IBM Quantum Experience}. This will include a discussion about quantum simulation without noise, with (mostly measurement) noise and treatment about noise mitigation. The experiments are performed in the quantum simulator \texttt{qasm}\_\texttt{simulator} and we use the device data from the quantum device \texttt{ibmq}\_\texttt{armonk} in April 2020. We use compact encoding for $\Delta_{\max}=9$ and we encode it into a three-qubit matrix. We use \texttt{qiskit} as our programming language. For a more technical introduction about \texttt{IBM Quantum Experience}, see the tutorial about \texttt{qiskit} in \cite{qiskit}.

\subsubsection{Without noise}
The following Figure \ref{fig2} is a simulation of VQE using \texttt{IBM Quantum Experience} without noise. We use the \texttt{VQE} function in \texttt{qiskit.aqua.algorithms.adaptive}. To extract our quantum simulation data in different measurements, we define a callback function to extract \texttt{eval\_count} and \texttt{mean} in \texttt{VQE}. We set our optimizer to be 
\begin{align}
\texttt{optimizer = SPSA(max\_trials=1000)}~,
\end{align}
where \texttt{SPSA} (Simultaneous Perturbation Stochastic Approximation) is an optimizer from \texttt{qiskit.aqua.components.optimizers} using \cite{kandala2017hardware}. Moreover, we use the variational ansatz
\begin{align}
\texttt{var\_form = RY(depth=3, entanglement='linear')}~.
\end{align}
The initial energy will quickly grow because of random initial steps, but it will decay quickly, obtaining a good convergence. Finally, we get the energy 5.89612083069727, which has a relative error $0.25\%$ from the exact result. 
\begin{figure}[H]
  \centering
  \includegraphics[width=1.0\textwidth]{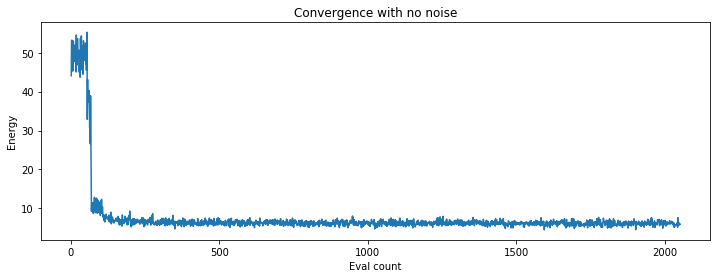}
  \caption{\label{fig2} VQE in \texttt{IBM Quantum Experience} without noise. We see that \texttt{mean} will quickly converge after several measurements. We get 5.89612083069727 after around 2000 evaluations, with the relative error $0.25\%$.}
\end{figure}

\subsubsection{Noisy quantum circuit}
\begin{figure}[H]
  \centering
  \includegraphics[width=1.0\textwidth]{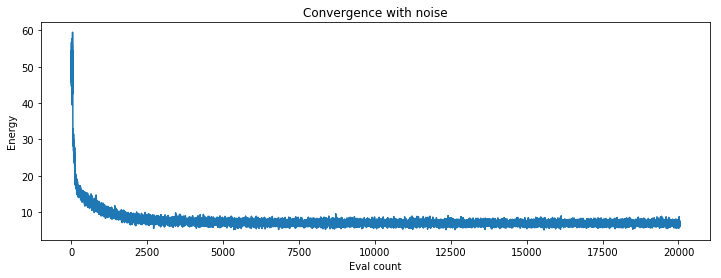}
  \caption{\label{fig3} VQE in \texttt{IBM Quantum Experience} with noise. Although we still get convergence, the result is much worse than before due to noise. We get the \texttt{mean} energy 7.129123262609374 energy after 20000 measurements. }
\end{figure}

Now we perform a similar experiment in Figure \ref{fig3} but with a noisy quantum circuit. We set the device to be the \texttt{IBM} quantum computer 
\begin{align}
\texttt{device = provider.get\_backend('ibmq\_armonk')}~,
\end{align}
to get the noise model
\begin{align}
\texttt{noise\_model = noise.device.basic\_device\_noise\_model(device.properties())}~.
\end{align}
But we still set the local backend 
\begin{align}
\texttt{backend = Aer.get\_backend('qasm\_simulator')}~,
\end{align}
for our local quantum simulation. Similarly, we use the optimizer
\begin{align}
\texttt{optimizer = SPSA(max\_trials=10000)}~,
\end{align}
and the variational ansatz
\begin{align}
\texttt{var\_form = RYRZ(depth=5, entanglement='linear')}~,
\end{align}
we could notice that the error is much larger due to noise in the measurement. In fact, after around 20000 measurements we get the \texttt{mean} energy 7.129123262609374, which is even larger than the CFT energy 6. Thus we need further error mitigation to reduce the error. 

\subsubsection{Error mitigation}
\begin{figure}[H]
  \centering
  \includegraphics[width=1.0\textwidth]{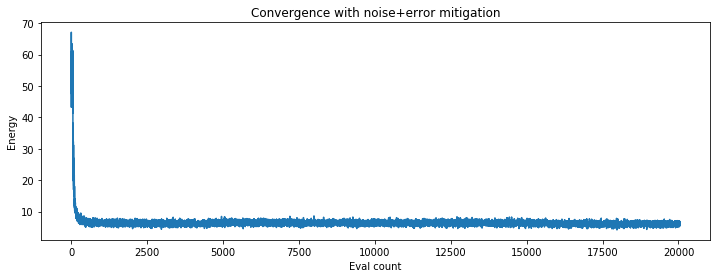}
  \caption{\label{fig4} VQE in \texttt{IBM Quantum Experience} with noise and error mitigation. After around 20000 measurements we get the \texttt{mean} energy 5.897078993668906 with relative error $0.26\%$.}
\end{figure}

Now we perform an experiment with the same noise model and further error mitigation (see Figure \ref{fig4}). We use \texttt{CompleteMeasFitter} from \texttt{qiskit.ignis.mitigation.measurement}. We run the simulator with the following \texttt{quantum\_instance},
\begin{align}
&\texttt{quantum\_instance = QuantumInstance}\nonumber\\
&\texttt{(measurement\_error\_mitigation\_cls=CompleteMeasFitter,} \nonumber\\
&\texttt{cals\_matrix\_refresh\_period=30)}~,
\end{align}
and we set the same optimizer and the same variational ansatz. Now we get much better computational performance with significantly less error. Finally, after around 20000 measurements we get the \texttt{mean} energy 5.897078993668906 with relative error $0.26\%$. We see that it is pretty similar to the result without noise. The above simulations show that near-term quantum computing for small conformal truncation problems is possible, and we do expect it will happen in some real near-term quantum devices.

\section{Other near-term algorithms}\label{other}
\subsection{Imaginary time evolution (ITE)}
Here we briefly describe another approach: imaginary time evolution (ITE). The idea is from a simple fact of quantum mechanics: if we evolve the Hamiltonian with imaginary time for a sufficiently long time, one projects the initial state to the ground state of the Hamiltonian.
 
There are mainly two types of imaginary time evolution algorithms existing in the current quantum information science. One type of algorithm is based on the locality of the Hamiltonian: assuming locality of the Hamiltonian and small time interval, one could use Lie-Trotter formula to expand the original Hamiltonian $H$ to small pieces $h_i$, where each $h_i$ only acts on a few qubits. Then, one could try to replace the imaginary time evolution by a unitary operator. The unitary operator is determined by quantum state tomography and solving linear equations \cite{sun}. Since in the conformal truncation case, there is usually no notion of lattice and locality, this method is not applicable generically in our situation. 
 
Thus we will describe another method \cite{ITEva}, which is based on the variational principle. Consider the form of the imaginary time evolution
\begin{align}
|\psi (\tau )\rangle  = A(\tau ){e^{ - H\tau }}|\psi (0)\rangle ~,
\end{align}
where
\begin{align}
A(\tau ) = \frac{1}{{\sqrt {\left\langle {\psi (0)\left| {{e^{ - 2H\tau }}} \right|\psi (0)} \right\rangle } }}~.
\end{align}
The Wick-rotated Schordinger equation is 
\begin{align}
\frac{\partial|\psi(\tau)\rangle}{\partial \tau}=-\left(H-E_{\tau}\right)|\psi(\tau)\rangle~,
\end{align}
where $E_{\tau}=\langle\psi(\tau)|H| \psi(\tau)\rangle$ is a constant from normalization. Again, we choose states in the space of ansatz $V$:
\begin{align}
\left| \psi (\vec{\theta})  \right\rangle  = {U_q}({\theta _q}) \ldots {U_2}({\theta _2}){U_1}({\theta _1})\left| {{\psi _{{\rm{simple}}}}} \right\rangle ~.
\end{align}
Using the McLachlan's variational principle:
\begin{align}
\delta \|\left(\partial / \partial \tau+H-E_{\tau}\right)|\psi(\tau)\rangle \|= 0~,
\end{align}
one could obtain a differential equation for the vector $\vec{\theta}$,
\begin{align}
\sum_{j} A_{i j} \dot{\theta}_{j}=C_{i}~,
\end{align}
where
\begin{align}
&{A_{ij}} = {\mathop{\rm Re}\nolimits} \left( {\frac{{\partial \langle \psi (\tau )|}}{{\partial {\theta _i}}}\frac{{\partial |\psi (\tau )\rangle }}{{\partial {\theta _j}}}} \right)~,\nonumber\\
&{C_i} = {\mathop{\rm Re}\nolimits} \left( { - \frac{{\partial \langle \psi (\tau )|}}{{\partial {\theta _i}}}H|\psi (\tau )\rangle } \right)~.
\end{align}
The coefficient $A_{ij}$ and $C_i$ could be measured by quantum circuits \cite{ITEva}. By measuring the parameters $A$ and $C$ at time $\tau$, one can formulate an update rule based on the differential equation. For instance, we could use the Eulerian method
\begin{align}
\vec{\theta}(\tau+\delta \tau) \simeq \vec{\theta}(\tau)+\dot{\vec{\theta}}(\tau) \delta \tau=\vec{\theta}(\tau)+A^{-1}(\tau) \cdot \vec{C}(\tau) \delta \tau~.
\end{align}
After evolving for a sufficiently long time, we could obtain a reasonable ground state.

\subsection{Quantum Lanczos algorithm (QLA)}
\subsubsection{Lanczos approach and its efficiency in conformal truncation}
Firstly we will review the classical Lanczos algorithm here, which is proven to be very efficient, usually in several diagonalization problems. 
 
Suppose that we have a Hamiltonian $H$, and we wish to find its ground state $\ket{\Omega}$ by variational approach. We know that the function $E(\ket{\psi})$ is minimized at $\ket{\psi}=\ket{\Omega}$ with the ground state energy $E_0$. The vector 
\begin{align}
\frac{{\delta E[\left| \psi  \right\rangle ]}}{{\delta \langle \psi |}} \equiv \frac{{H\left| \psi  \right\rangle  - E[\left| \psi  \right\rangle ]\left| \psi  \right\rangle }}{{\langle \psi |\psi \rangle }} = \left| {{\psi _a}} \right\rangle ~,
\end{align}
gives the direction towards the actual ground state with the steepest ascent. Thus we move in the direction $\ket{\psi}-\gamma \ket{\psi_a}$ for some small $\gamma>0$. 
 
The idea of Lanczos algorithm is trying to find the optimal $\gamma$, which means that we wish to minimize $E\left[ {\ket{\psi}  - \gamma {\ket{\psi_a}}} \right]$. Namely, we need to search for an optimal vector in the space
\begin{align}
K = {\mathop{\rm span}\nolimits} \left( {|\psi \rangle ,\left| {{\psi _a}} \right\rangle } \right) = {\mathop{\rm span}\nolimits} (|\psi \rangle ,H|\psi \rangle )~.
\end{align}
One way to find the optimal vector is to find the state with minimal energy in $K$. By constructing the basis in $K$, one could construct an effective Hamiltonian associated with the space $K$ and find the ground state. 
 
The above algorithm could be iterated several times. We define the $(L+1)$-dimensional, so-called \emph{Krylov space}
\begin{align}
{K^L}\left( {\left| {{v_0}} \right\rangle } \right) = {\mathop{\rm span}\nolimits} \left( {\left| {{v_0}} \right\rangle ,H\left| {{v_0}} \right\rangle ,{H^2}\left| {{v_0}} \right\rangle , \ldots ,{H^L}\left| {{v_0}} \right\rangle } \right)~,
\end{align}
where $\ket{v_0}$ is an initial normalized state with non-zero overlap with the ground state. The basis of $K^L$ could be defined iteratively. We define
\begin{align}
{a_j} = \left\langle {{v_j}|H|{v_j}} \right\rangle ~,
\end{align}
where $\ket{v_j}$ is a set of orthonormal states.  We construct
\begin{align}
&{b_1}\left| {{v_1}} \right\rangle  = \left| {{{\tilde v}_1}} \right\rangle  = H\left| {{v_0}} \right\rangle  - {a_0}\left| {{v_0}} \right\rangle~, \nonumber\\
&{b_{j + 1}}\left| {{v_{j+ 1}}} \right\rangle  = H\left| {{v_j}} \right\rangle  - {a_j}\left| {{v_j}} \right\rangle  - {b_j}\left| {{v_{j - 1}}} \right\rangle ~,
\end{align}
where $b_{j}^{2}=\left\langle\tilde{v}_{j} | \tilde{v}_{j}\right\rangle$. Then, one could compute an effective Hamiltonian
\begin{align}
{H_{{K^L}\left( {\left| {{v_0}} \right\rangle } \right)}} = \left( {\begin{array}{*{20}{c}}
{{a_0}}&{{b_1}}&0&0&0&0&{}\\
{{b_1}}&{{a_1}}&{{b_2}}&0& \cdots &0&0\\
0&{{b_2}}&{{a_2}}&{{b_3}}&{}&0&0\\
0&0&{{b_3}}&{{a_3}}&{}&0&0\\
{}& \vdots &{}&{}& \ddots & \vdots &{}\\
0&0&0&0&{}&{{a_{L - 1}}}&{{b_L}}\\
0&0&0&0& \cdots &{{b_L}}&{{a_L}}
\end{array}} \right)~.
\end{align}
After diagonalizing the Hamiltonian, one could compute an approximate ground state 
\begin{align}
\left| \Omega  \right\rangle  \approx \sum\limits_{j = 0}^L {{{\tilde \psi }_{0,j}}\left| {{v_j}} \right\rangle } ~,
\end{align}
where $\tilde{\psi}_{0}$ is the ground state vector of the effective Hamiltinian ${H_{{K^L}\left( {\left| {{v_0}} \right\rangle } \right)}}$. One could also estimate the $n$-th low-lying excited state $\Omega_n$ as
\begin{align}
\left| {{\Omega _n}} \right\rangle  \approx \sum\limits_{j = 0}^L {{{\tilde \psi }_{n,j}}\left| {{v_j}} \right\rangle } ~,
\end{align}
where $\tilde{\psi}_{n}$ is the eigenvector of ${H_{{K^L}\left( {\left| {{v_0}} \right\rangle } \right)}}$ in the $n$-th low-lying eigenspace.  

As we mention before, sparsity is a generic feature for a large class of conformal truncation problems. Lanczos-type methods are particularly useful when the Hamiltonian is sparse. The 2D QCD Hamiltonian is a little special because we keep only the two-particle Hilbert space. For finite $N_c$ QCD, which still has particle number changing matrix elements, the Hilbert space is much larger, and the Hamiltonian is even much more sparse, and we would expect Lanczos to give a significant speed-up. In other works \cite{Fitzpatrick:2019cif,Anand:2017yij}, the Hamiltonian is also very sparse, and the Lanczos-type methods play an important role in diagonalizing their Hamiltonian.

\subsubsection{Quantum implementation}
Now let us describe a quantum operational framework of the Lanczos approach, which is called the Quantum Lanczos algorithm (QLA). In classical construction, the development of $\ket{v_n}$ is ensured to be orthogonal. This will provide simplification and fast implementation in the classical devices, but may not be very practical for quantum devices, especially in the near term. Thus, we could think about formulating an arbitrary basis of the Krylov space, which may not be necessary to be orthogonal.

Ideally, suppose that we could prepare the states $ {\left| {{v_0}} \right\rangle ,H\left| {{v_0}} \right\rangle ,{H^2}\left| {{v_0}} \right\rangle , \ldots ,{H^L}\left| {{v_0}} \right\rangle }$ efficiently in a quantum computer, one could take those states themselves as the basis. There are relatively simpler tricks to realize this computation, which is a combination of \cite{sun} and the Quantum Subspace Expansion method developed in \cite{QSM1,QSM2}: one could construct $L+1$ states,
\begin{align}
\left| {{v_l}} \right\rangle  = \frac{{{H^l}\left| {{v_0}} \right\rangle }}{{\left\| {{H^l}\left| {{v_0}} \right\rangle } \right\|}} \equiv {n_l}{H^l}\left| {{v_0}} \right\rangle~,
\end{align}
for $l=0,1,\cdots,L$. We could define
\begin{align}
S_{l, l^{\prime}}=\left\langle v_{l} |  v_{l^{\prime}}\right\rangle \quad ~,~~~~~ \quad H_{l, l^{\prime}}=\left\langle v_{l}|{H}|  v_{l^{\prime}}\right\rangle~,
\end{align}
and we call the corresponding matrices $\mathbf{S}$ and $\mathbf{H}$ in the Krylov space\footnote{There is a trick to evaluate those matrices in a quantum device. For both even or odd indices, defining $2r=l+l'$, we have
\begin{align}
&{S_{l,l'}} = \frac{{{n_l}{n_{l'}}}}{{n_r^2}}~,\nonumber\\
&{H_{l,l'}} = {S_{l,l'}}\left\langle {{v_r}|H|{v_r}} \right\rangle ~.
\end{align}
Moreover, one could estimate $n_r$ iteratively
\begin{align}
\frac{1}{{n_{r + 1}^2}} = \frac{{\left\langle {{v_r}\left| {{H^2}} \right|{v_r}} \right\rangle }}{{n_r^2}}~.
\end{align}
Using iterations, we could fix all $n_l$s and then we could compute $\mathbf{S}$ and $\mathbf{H}$ directly.}. Then, one could solve the generalized eigenvalue problem
\begin{align}
{\bf{Hx}} = E{\bf{Sx}}~,
\end{align}
and then one could get $n$-th low lying eigenstates as  
\begin{align}
\left| {{\Omega _n}} \right\rangle  \approx \sum\limits_l {x_l^n} \left| {{v_l}} \right\rangle ~,
\end{align}
where $x_l^n$ is the eigenvector in the $n$-th eigenspace.
\subsubsection{Variational-based QLA}
The above algorithm is theoretically explicit but might be hard to perform in the quantum devices. One of the main problems is iteratively applying a non-local Hamiltonian $H$, which may not be very easy for near term devices. Thus, we might combine the variational technics, Lanczos algorithm, and imaginary time evolution to get a hybrid approach for near term devices. 
 
Firstly, we wish to replace the original Krylov space by \cite{sun}
\begin{align}
{\tilde{K}^L}\left( {\left| {{v_0}} \right\rangle } \right) = {\mathop{\rm span}\nolimits} \left( {\left| {{v_0}} \right\rangle ,{e^{ - \Delta \tau H}}\left| {{v_0}} \right\rangle ,{e^{ - 2\Delta \tau H}}\left| {{v_0}} \right\rangle , \ldots ,{e^{ - \Delta \tau HL}}\left| {{v_0}} \right\rangle } \right)~,
\end{align}
which is straightforward in the quantum implementation. In this case (imaginary time evolution), we are easier to get access to the low energy subspace. Unlike \cite{sun}, in order to solve the imaginary time evolution, we will use the variational method that is discussed before. Thus, we obtain a variational-based quantum Lanczos algorithm, that is in principle, implementable in the near-term devices. 
\subsection{Trial numerics}
In this section, we provide some trials for those near-term algorithms. We will focus on similar setups to our previous VQE calculations. For a given fixed $\Delta_{\max}$, we drop out the massless pion contribution and focus on the first excited meson, which is the ground state of the effective Hamiltonian without the massless pion.   

\subsubsection{ITE}
\begin{figure}[H]
  \centering
  \includegraphics[width=1.0\textwidth]{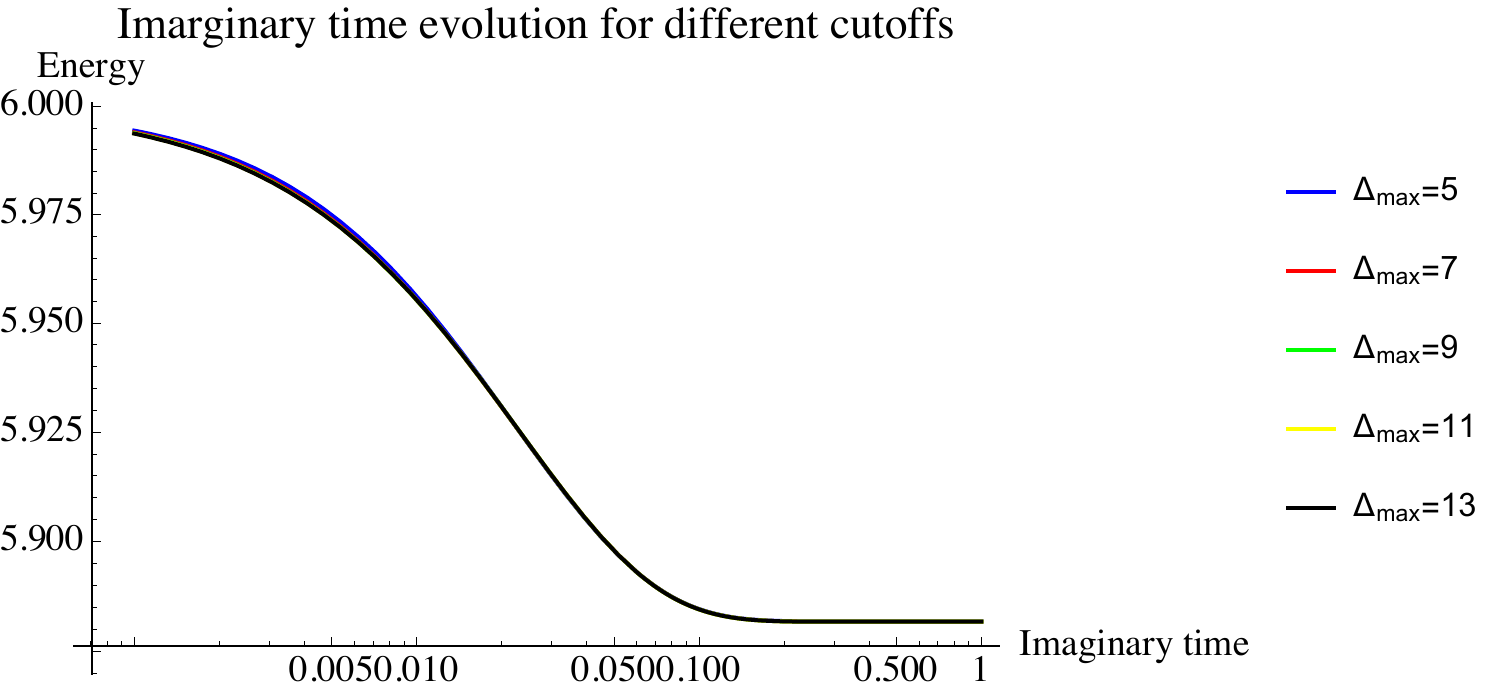}
  \caption{\label{fig5} ITE calculation for different cutoffs. We find that for different $\Delta_{\max}$s using ITE, we could obtain very similar convergence.}
\end{figure}

\begin{figure}[H]
  \centering
  \includegraphics[width=1.0\textwidth]{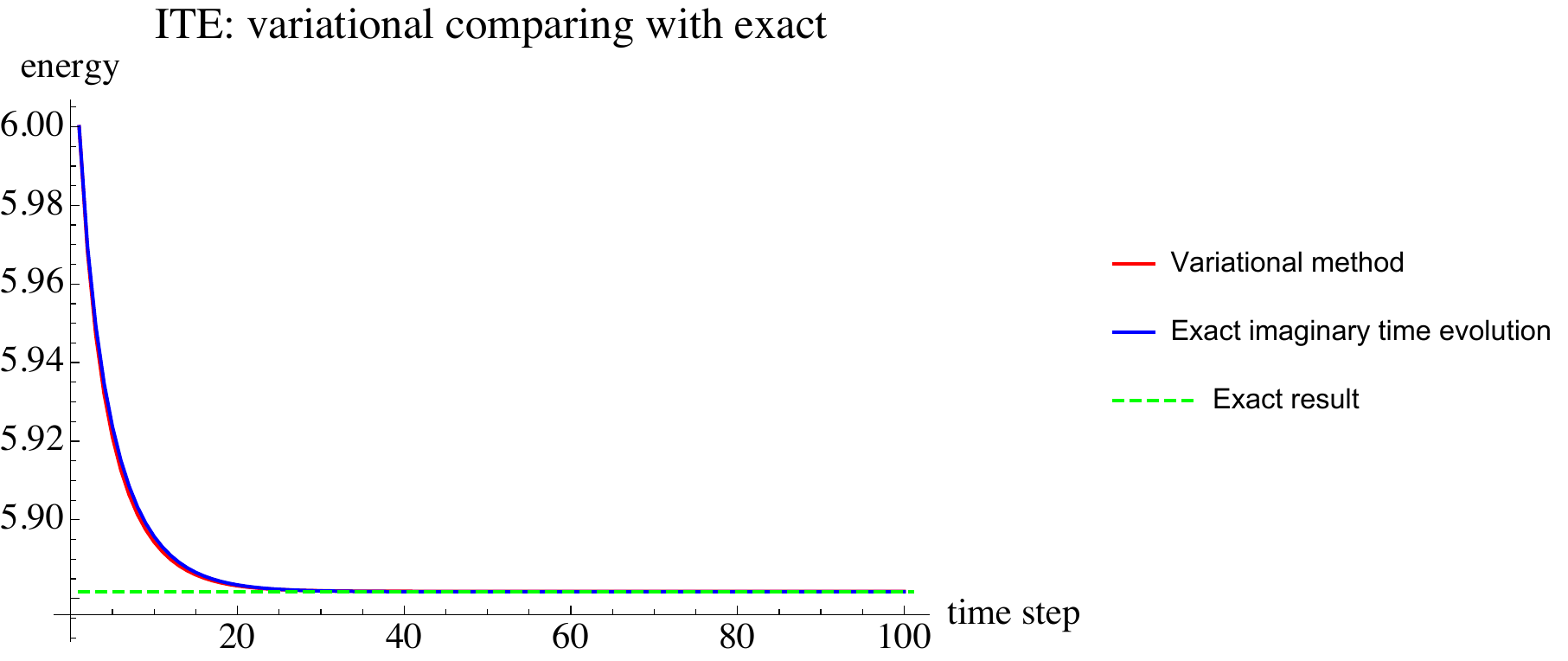}
  \caption{\label{fig6} ITE comparison between variational and exact ITE. For the variational calculation, we run 100 steps by setting the gradient descent momentum $\gamma=0.006$. We use the same variational form as the VQE calculation: the product of rotating Pauli $Y$s. }
\end{figure}

\begin{figure}[H]
  \centering
  \includegraphics[width=0.7\textwidth]{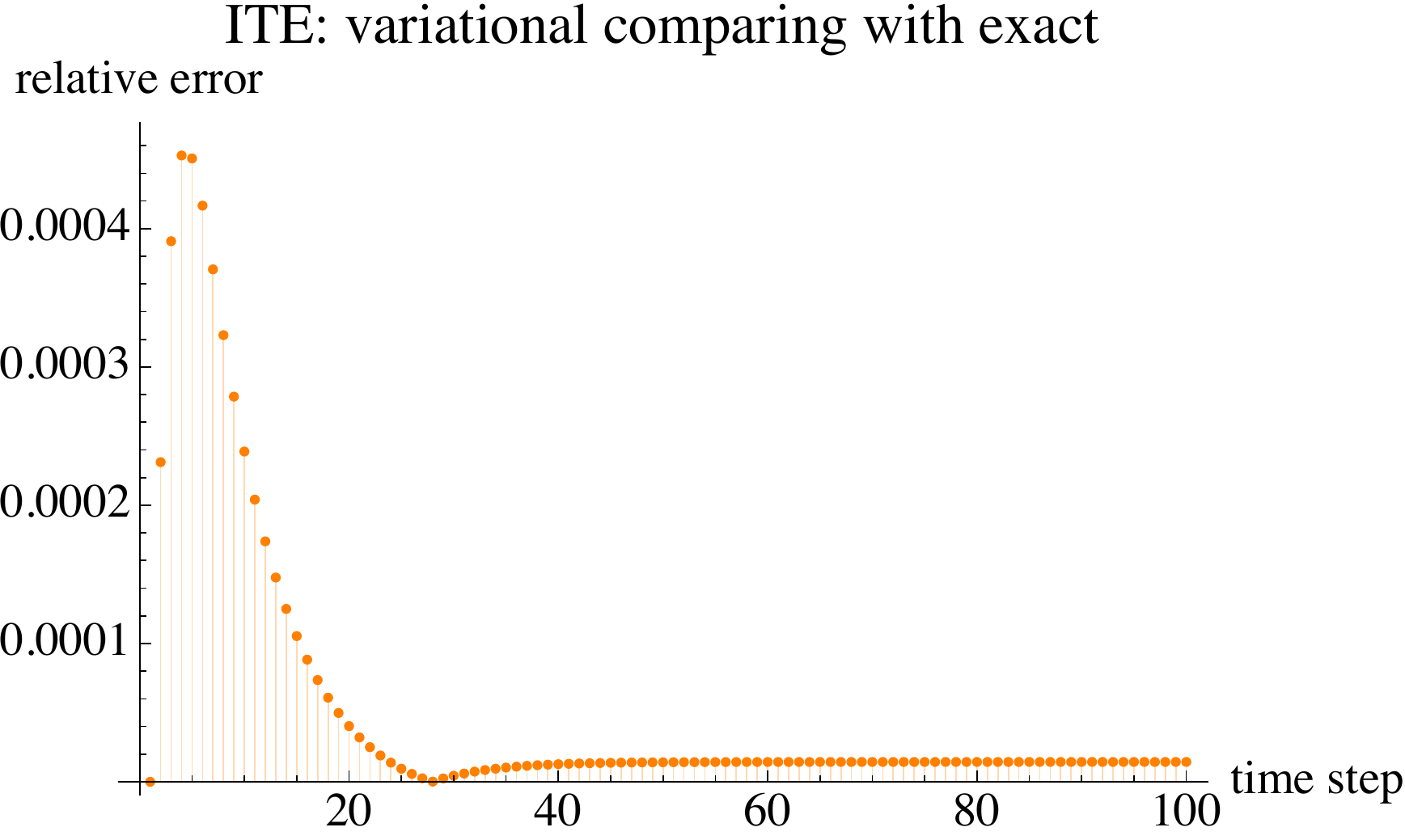}
  \caption{\label{fig7} The relative error between variational and exact ITE. For the variational calculation, we run 100 steps by setting the gradient descent momentum $\gamma=0.006$. We use the same variational form as the VQE calculation: the product of rotating Pauli $Y$s.}
\end{figure}

The first numerical trial we wish to present is the convergence performance of ITE for different choices of $\Delta_{\max}$ (see Figure \ref{fig5}). We start from the initial state as the CFT energy eigenstate with energy 6. We find that for different $\Delta_{\max}$s, we get very similar convergence. For a well-behaved conformal truncation problem, we should achieve such stability after some certain $\Delta_{\max}$s, and we expect this story is generic. 
Since we have a variational approach for ITE that is more suitable for near-term devices, we make a simple comparison between variational and exact versions of ITE. Figure \ref{fig6} and Figure \ref{fig7} show the promised comparison. We see that in our variational algorithm, ITE behaves pretty well, and we obtain a relatively small error. 

\subsubsection{QLA}
\begin{figure}[H]
  \centering
  \includegraphics[width=0.7\textwidth]{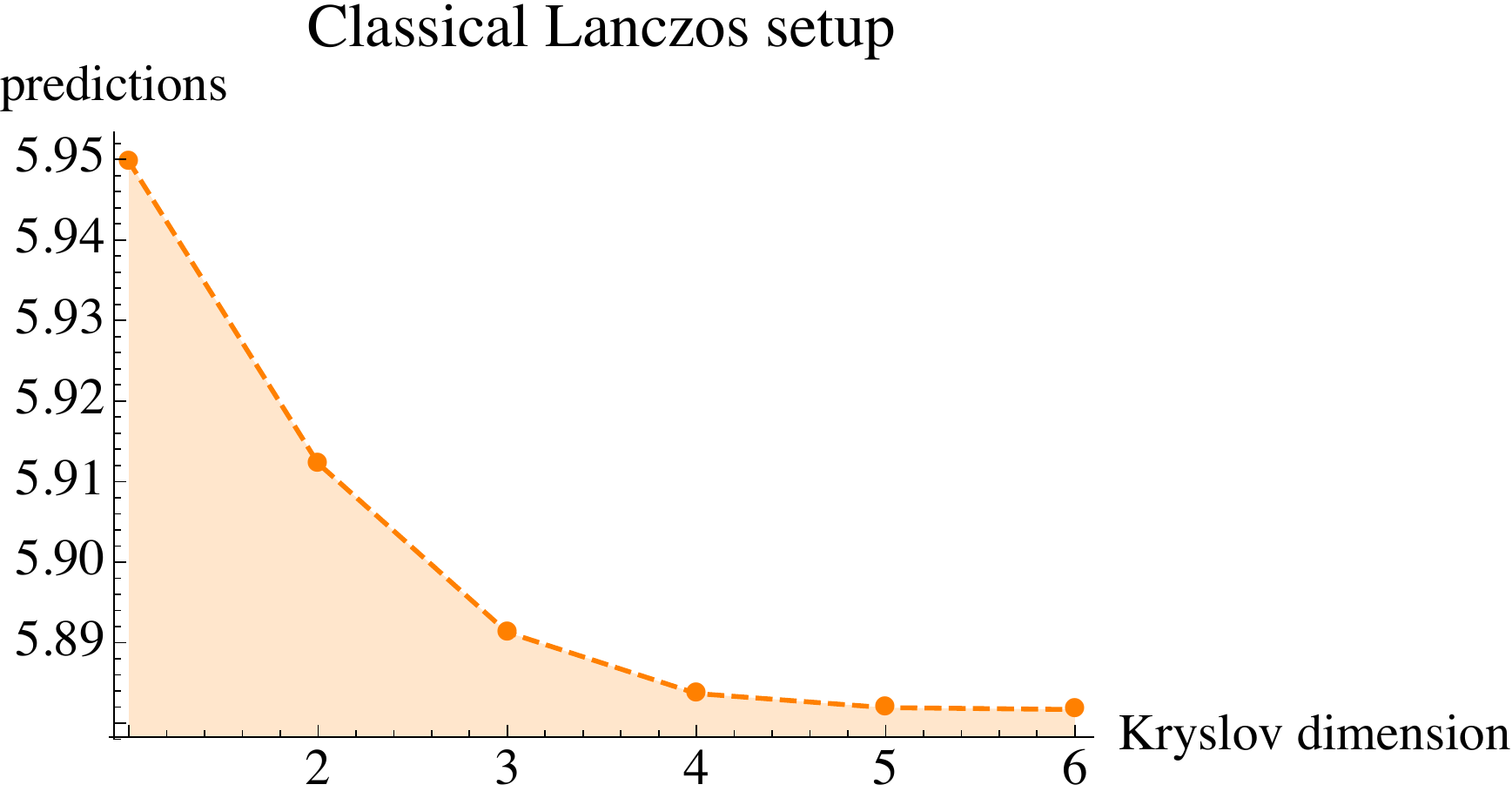}
  \caption{\label{fig8} Numerical simulation of classical Lanczos algorithm using different Krylov space dimensions. We take $\Delta_{\max}=16$ in these calculations.}
\end{figure}

\begin{figure}[H]
  \centering
  \includegraphics[width=0.7\textwidth]{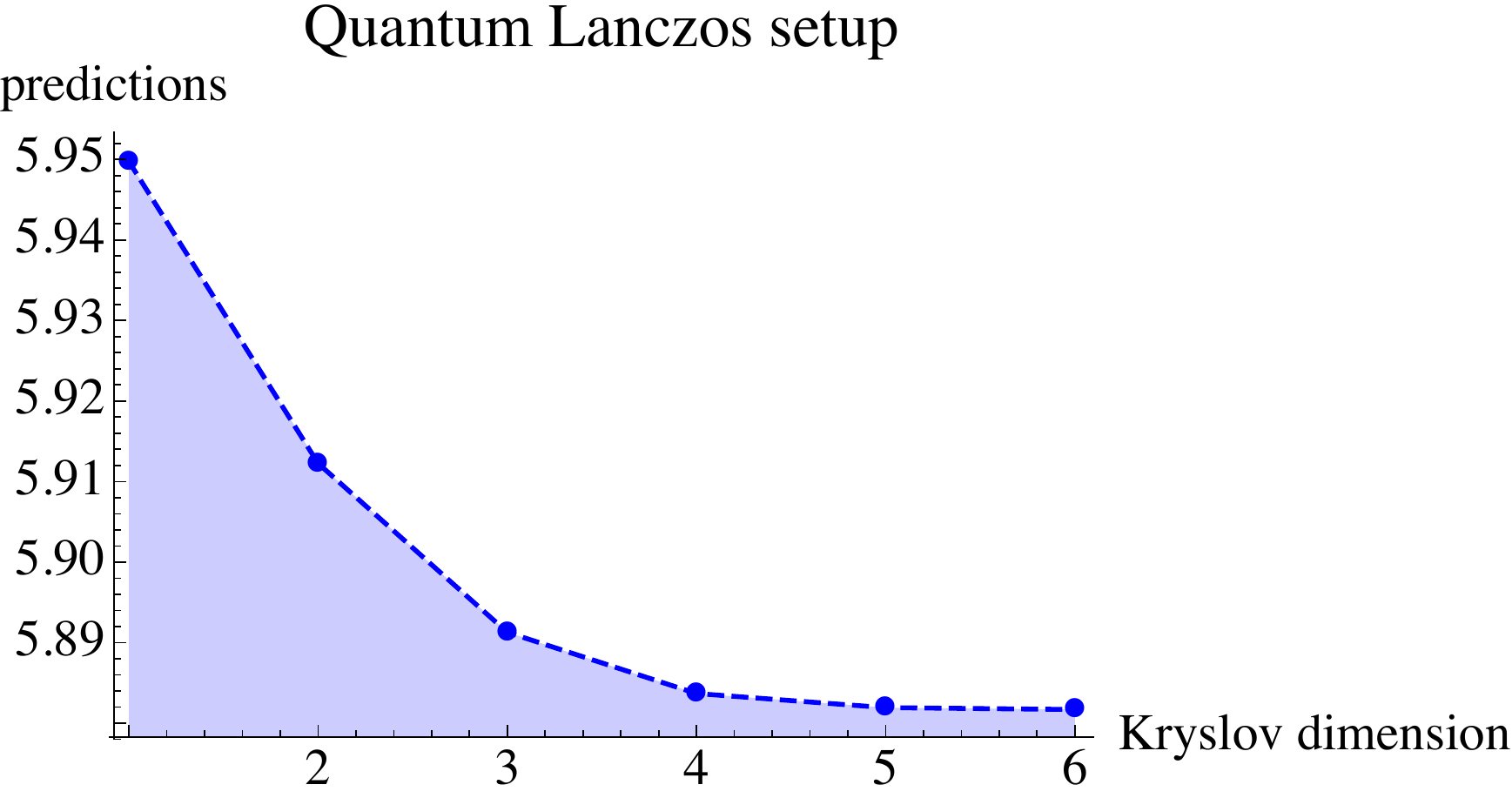}
  \caption{\label{fig9} Numerical simulation of QLA using different Krylov space dimensions. We take $\Delta_{\max}=16$ in these calculations. Note that in this calculation we directly diagonalize the space $\text{span}(H^k\left|v_{0}\right\rangle)$, not its ITE version. }
\end{figure}

\begin{figure}[H]
  \centering
  \includegraphics[width=1.0\textwidth]{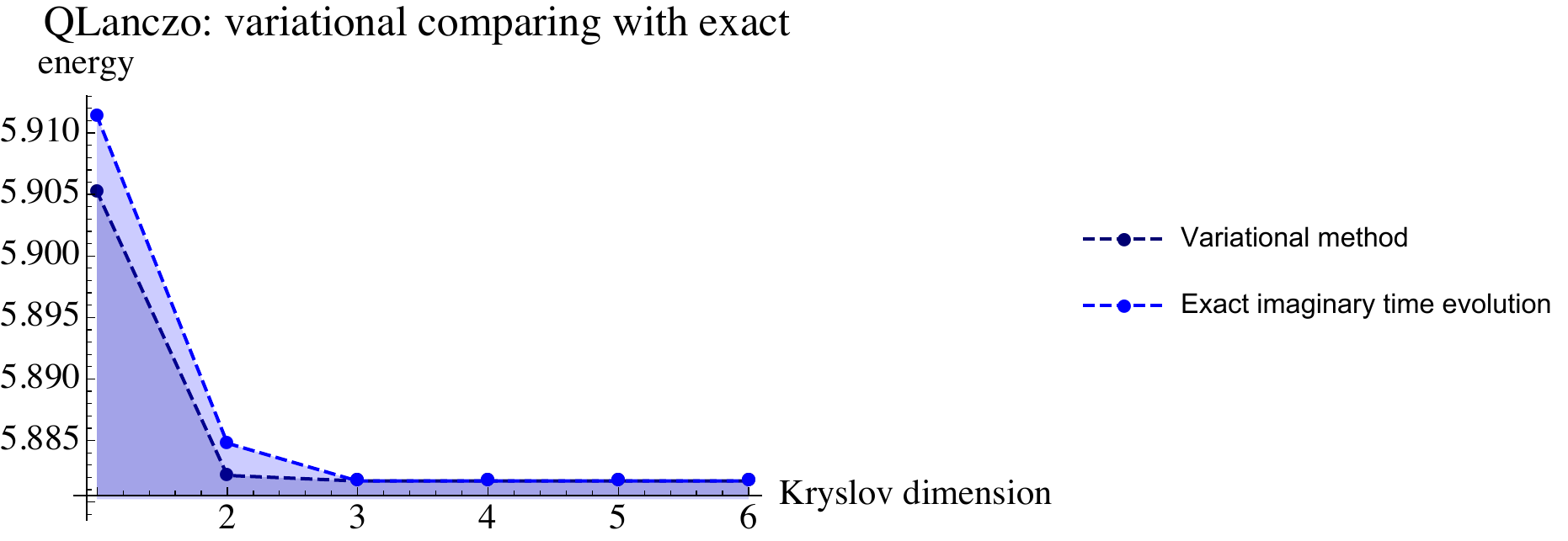}
  \caption{\label{fig10} A comparison between variational and exact QLA in their ITE versions. Now we take $\gamma=0.006$, and we only run for 7 steps. We see that the variational version of QLA even behaves better than the exact version in this toy example.}
\end{figure}

\begin{figure}[H]
  \centering
  \includegraphics[width=0.7\textwidth]{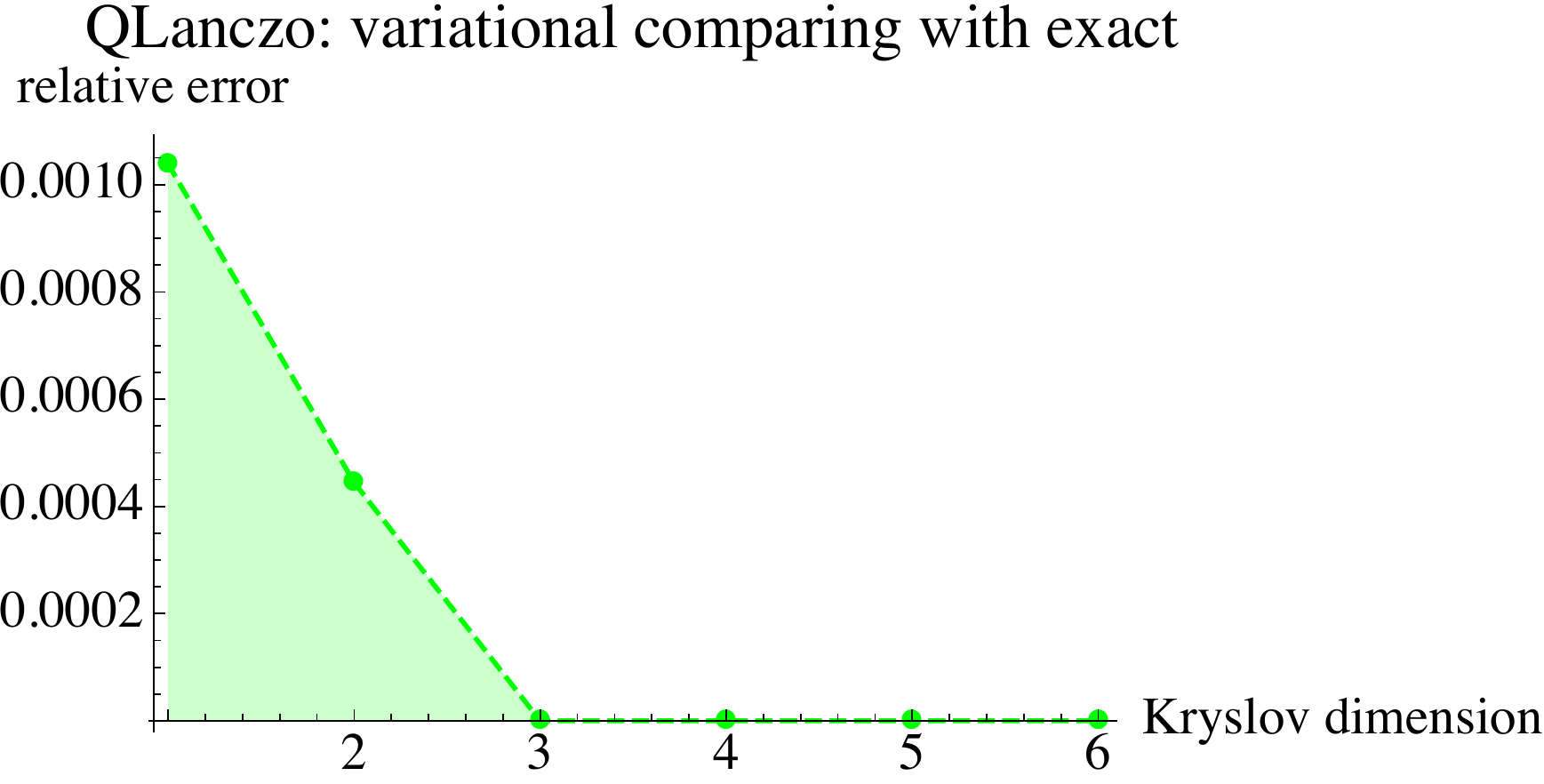}
  \caption{\label{fig11} The relative error between variational and exact QLA in their ITE versions. Now we take $\gamma=0.006$ and we only run for 7 steps. The relative error is shown to be very small.}
\end{figure}

We also make some numerical implementations of QLA. Here, we mainly look at the convergence performance for different dimensions of Krylov spaces. We perform the classical Lanczos algorithm with $\Delta_{\max}=16$ in Figure \ref{fig8}. We also perform its quantum counterpart (using the space $\text{span}(H^k\left|v_{0}\right\rangle)$) with $\Delta_{\max}=16$ in Figure \ref{fig9}. Considering that the primary difference between classical and quantum Lanczos algorithm in our current simulation is that in the classical implementation, we construct a new sparse matrix. Since the original Hamiltonian is very sparse (which is claimed before to be generic in several Hamiltonian truncation problems), we would expect that the quantum version is not that different from the classical version at this stage in terms of errors.
Furthermore, we uplift our QLAs towards their finite temperature (ITE) version and their variational version and make a comparison in Figure \ref{fig10} and Figure \ref{fig11} with $\Delta_{\max}=9$. We see that the variational one even behaves better than the exact one in this toy example, and the relative error is very small. 
The above simple trials show that quantum implementations of those near-term algorithms in the real devices are also possible.

\section{Conclusions and opportunities}\label{conc}
In this work, we initialize the study of generic quantum field theories using the quantum simulation of conformal truncation. We design quantum algorithms that are, in principle, applicable for near-term and future quantum computers for quantum simulation of conformal truncation. Conceptually, the physics of the conformal truncation method reveals a novel way different from the current mainstream approach of quantum simulation of quantum field theories (for instance, quantum algorithms based on Jordan-Lee-Preskll \cite{Jordan:2011ne,Jordan:2011ci}). Moreover, it has a manifest description of RG in the Hilbert space language. Our work shows that the conformal truncation approach might provide a valuable improvement on the current and future research of quantum simulation of quantum field theories, complementary to the existing lattice approach.
 
Here, we summarize some possible future plans along this line. 

\subsection{\texttt{Qiskit.qft}}
This work opens a novel direction of solving quantum field theory in quantum devices using a novel way. In the future, we should try to improve the current calculation of conformal truncation, especially for higher dimensions and more non-trivial UV fixed points, with the help of near-term or universal quantum devices. We are not sure when this could happen with a \emph{quantum supremacy} computation in the quantum field theory context, but we believe that it might eventually happen in the future. In the near-term, works could be done for more examples. As long as we know an explicit form of a Hamiltonian for a given conformal truncation problem, we could import it in a quantum setting. Thus, we call it \texttt{Qiskit.qft}, a potential package for a dictionary of quantum field theories, as an analog of \texttt{Qiskit.chemistry}, an existing package in the \texttt{IBM quantum experience}. 

\subsection{Quantum Church-Turing Thesis}
It is also interesting to study conceptual issues about the quantum Church-Turing Thesis. Conformal truncation provides a novel perspective towards previous arguments about the quantum Church-Turing Thesis, and maybe we could have a conceptually-improved understanding of simulating strongly-coupled quantum field theories. Furthermore, recently there are some important progress about the quantum Church-Turing Thesis in curved spacetime and black holes \cite{Bouland:2019pvu,Susskind:2020kti,Kim:2020cds}. It will be interesting to study those proposals in field theories.

\subsection{Renormalization}
We describe some conceptual interpretations of renormalization in the quantum simulation of conformal truncation problems. It will be interesting to connect other proposals about renormalization, quantum simulation, and variational calculations or dive deeper into the theoretical formalism of understanding variational simulation as an RG flow in general. It might also be interesting to study RG flows in the quantum information language \cite{toappear}.

\subsection{Fair comparison}
As we mentioned before, for us, it is not very clear which method is better comparing the lattice method and the conformal truncation. A possible situation is that those methods have their own advantages in different computational tasks. However, we think it will be very important to think about a potential, detailed comparison between them. A fair comparison means that with comparable computational resources, people should show that one method has better computational performance than the other in the same field theory task. It is also important to notice that since the lattice method and the conformal truncation approach have different starting points: one is from the free theory while the other is from the UV fixed point, there are extra theoretical works remaining for clarifying how to match bared and physical parameters on each side.

\subsection{Possible quantum advantage and benchmark}
Our design of digital quantum simulation is based on adiabatic state preparation and adiabatic quantum computing, which is well-known for the possibility of solving hard problems in the polynomial time. Thus, it might be interesting to give some concrete estimations, both in the abstract and practical sense, on precise computational cost due to adiabatic quantum computation. In principle, a precise estimate of computational cost will require detailed information about the corresponding quantum devices in mind, and the quantum field theory itself we wish to explore. For near-term devices, it also might be useful to provide meaningful physics problems to test and benchmark the corresponding machines, for instance, the scaling of the computational cost and the computational error.  

\subsection{Large-scale bootstrap}
Since the conformal approach requires a full set of conformal data, it is very hard to study conformal truncation problems from non-trivial UV fixed points since we usually don't have full control of the conformal data. This provides a natural motivation for conformal bootstrap. Recent developments in large-scale bootstrap problems based on new implementation of semidefinite programming solver \cite{ElShowk:2012ht,Simmons-Duffin:2015qma,Landry:2019qug,Chester:2019ifh} might provide potential capability for clearer resolution of conformal spectra in non-trivial CFTs especially in higher dimensions \cite{Simmons-Duffin:2016wlq}. Thus, we would expect that with updated technologies in conformal bootstrap, we are able to obtain better results of conformal truncation using both classical and quantum devices. Moreover, conceptually it might be interesting to connect quantum convex programming algorithms and conformal bootstrap physics \cite{bao2019quantum}.

\subsection{Real-time experiments}
In this paper, although we use adiabatic state preparation, we are staying in regimes of static spectra. However, it will be interesting to explore further issues about real-time dynamics. Since we already know the Hamiltonian, similar algorithms we use for state preparation will be helpful for exploring time evolution, although we need to make a more detailed analysis. The particularly important problem is that since we truncate the Hamiltonian, the estimates of low-lying states are more precise than highly-excited states generically. Thus, we need to be careful about the validity and potential errors of real-time experiments for initial high-energy particles.

\subsection{Ideas about analog quantum simulation}
Finally, we make a few comments about analog quantum simulation. This paper only discusses digital and variational simulations, but analog simulation is also exciting to explore. In the analog simulation, we need to consider encoding our quantum states into some physical models made by, for instance, Rydberg atoms. It is also interesting to study how to map our Hamiltonians to some specific quantum chemistry models or spin-chain models as well. In high energy physics, people have studied some spin-chain interpretations of the dilatation generator in models such as $\mathcal{N}=4$ super Yang-Mills theory (for instance, see \cite{Beisert:2003yb}). It might be interesting to explore this type of interpretation beyond integrable models and try to connect with analog quantum simulation\footnote{We thank David Simmons-Duffin for a related discussion.}.

\section*{Acknowledgement}
We thank Fernando Brandao, Alex Buser, Cliff Cheung, Liam Fitzpatrick, Hrant Gharibyan, Alexandru Gheorghiu, Masanori Hanada, Masazumi Honda, Emanuel Katz, Isaac Kim, Henry Lamm, Peter Love, Ashley Milsted, John Preskill, Burak Sahinoglu, David Simmons-Duffin, Ning Su, Yuan Su, Chong Sun, Jinzhao Sun, Guifre Vidal, Xiao Yuan, Qi Zhao and You Zhou for related discussions. We thank Nikhil Anand, Liam Fitzpatrick, Emanuel Katz, Zuhair Khandker, Matthew Walters for letting us use some preliminary results in the LCT formulation of 2D QCD. We thank the Simons collaboration annual meetings \emph{It from qubit} and \emph{Non-perturbative bootstrap} for inspiring talks and exciting environment in 2019, where this work has been initiated. JL is supported in part by the Institute for Quantum Information and Matter (IQIM), an NSF Physics Frontiers Center (NSF Grant PHY-1125565) with support from the Gordon and Betty Moore Foundation (GBMF-2644), by the Walter Burke Institute for Theoretical Physics, and by Sandia Quantum Optimization \& Learning \& Simulation, DOE Award \#DE-NA0003525.

\appendix

\newcommand{\nn}{\nonumber}
\newcommand{\CP}{{\cal P}}
\newcommand{\kvec}{\boldsymbol{k}}

\section{Technical details of 2D QCD conformal truncation}\label{technical}
\label{sec:QCDappendix}

In this appendix, we show the details of computing the 2D QCD Hamiltonian matrix elements. The method described in this section is general enough to efficiently compute the matrix element for any $N_c$, although the text primarily focuses on the large $N_c$ limit, where the Hamiltonian is much simplified.

\subsection{Wick contraction}
In (\ref{eq:MatrixElements}) we introduce the matrix element as a Fourier transformation of the three-point correlation function
\begin{align}
\CM_{\CO\CO^\prime} = 
\int dxdydz \, e^{ipx - ip^\prime z}
\big\< \CO(x)\, 
(\psi^\dagger_i T_{ij}^A \psi_j) \frac{1}{\d^2} (\psi^\dagger_k T_{kl}^A \psi_l) (y)\, 
\CO^\prime(z) \big\> ~.
\end{align}
Recall the operators $\CO$ and $\CO^\prime$ in the external states are primary operators as sums of products $\d^k \psi$ and $\d^k \psi^\dagger$, which we call \emph{monomials} so that we can write
\begin{align}
\CO &\supset \d^{\kvec} \psi = \partial^{k_1} \psi_i^\dagger \partial^{k_2} \psi_i
\cdots ~,\nn \\
\CO^\prime &\supset \d^{\kvec^\prime} \psi = \partial^{k_1^\prime} \psi_j^\dagger \partial^{k_2^\prime} \psi_j \cdots ~,
\end{align}
and compute the building block matrix element between a pair of monomials 
\begin{equation}
\CM_{\CO\CO^\prime} \supset \CM_{\kvec\kvec^\prime}
\equiv \int dxdydz \, e^{iPx - iP^\prime z} \,
\big\< \d^{\kvec^\prime} \psi(x)\, 
(\psi^\dagger_i T_{ij}^A \psi_j) \frac{1}{\d^2} (\psi^\dagger_k T_{kl}^A \psi_l) (y)\, 
\d^{\kvec^\prime} \psi^\prime(z) \big\> ~.
\end{equation}
In the coordinate space, the nonlocal kernel $1/\d^2$ is an integral
\begin{align}
(\psi_i^\dagger T_{ij}^A \psi_j) \frac{1}{\d^2} (\psi_k^\dagger T_{kl}^A \psi_l) (y)
= (\psi_i^\dagger T_{ij}^A \psi_j)(y) \int^y dy^{\prime} \int^{y^\prime} dy^{\prime\prime} (\psi_k^\dagger T_{kl}^A \psi_l) (y^{\prime\prime}) ~.
\end{align}
To properly define the integrals in the coordinate space, one needs to specify the appropriate boundary condition at $y^\prime,y^{\prime\prime} \rightarrow \infty$. Instead of discussing it in coordinate space, we later go to momentum space, and the boundary condition becomes the principal value prescription. We use Wick contraction to compute more complicated matrix elements. We build up the spatial correlation functions out of two-fermion building blocks
\begin{equation}
\corr{\d^{k} \psi_i^\dagger(x) \d^{k'} \psi_j(y)} =\frac{\Gamma(k+k'+1)}{4\pi(x-y)^{k+k'+1}}  \delta_{ij} ~. \tag{\ref{eq:gen2ptfunctionQCD}  }
\end{equation}
We first hold the integral, and the compute the spacial four-point function $G(x,y,y^{\prime\prime},z)$ as an integrand, and for each contraction term we will obtain the following general form
\begin{align}\label{eq:gauge-5pt-func}
G(x, y, y^{\prime\prime}, z) = &
\frac{A_{\kvec,\kvec^\prime}}{
    (x-y)^a (y-z)^b (x-y^{\prime\prime})^{a^\prime}
    (y^{\prime\prime}-z)^{b^\prime}
    (x-z)^{c}
} ~,
\end{align} 
where $A_{\kvec,\kvec^\prime}$ is the product of $\Gamma$ and $4\pi$ factors from the two-point functions (\ref{eq:gen2ptfunctionQCD}).
Then we integrate out $y^{\prime\prime}$ and Fourier transform to get the matrix elements
\begin{align}\label{eq:gauge-matrix-element-compact}
\CM_{\kvec,\kvec^\prime} &\equiv \int dxdydz \, e^{iPx - iP^\prime z}
 \int^y dy^{\prime} \int^{y^\prime} dy^{\prime\prime} G(x, y, y^{\prime\prime}, z) \nn \\
 &= (2\pi) \delta(P-P^\prime)\frac{
    4\pi^2 i^{\Delta+\Delta^\prime-2}
    P^{\Delta+\Delta^\prime}
    A_{\kvec,\kvec^\prime}
}{\Gamma(\Delta+\Delta^\prime-1)} I(a,b,a^\prime,b^\prime) ~,
\end{align}
where we eliminate $c$ using the relation $a+b+a^\prime+b^\prime+c = \Delta+\Delta^\prime-2$ from dimensional analysis.
The integral is subject to IR divergence and is sensitive to the boundary condition. The correct treatment is equivalent to taking the self-energy shift, and the principal value integral in the momentum space, and the resulting factor $I(a,b,a^\prime,b^\prime)$ is nontrivial.
\subsection{General case}
We start from the general form 
\begin{align}\tag{\ref{eq:gauge-5pt-func}}
G(x, y, y^{\prime\prime}, z) = &
\frac{A_{\kvec,\kvec^\prime}}{
    (x-y)^a (y-z)^b (x-y^{\prime\prime})^{a^\prime}
    (y^{\prime\prime}-z)^{b^\prime}
    (x-z)^{c}
} ~.
\end{align}
Next, we make a Fourier transform for each spacial factor
\begin{align}
\int \frac{e^{i p x}\, dx}{(x-i \epsilon )^a} = \frac{2 \pi  i^a p^{a-1} \theta (p)}{(a-1)!} ~,
\end{align}
to expand the correlation function in parton momenta
\begin{align}
&\pr{\frac{\d}{\d y^{\prime\prime}}}^{-2}G(x, y, y^{\prime\prime}, z) \nn \\
=& A_{\kvec,\kvec^\prime}\int dp_1 dp_2 dp_1^\prime dp_2^\prime dq\, 
e^{-i x (p_1+p_1^\prime+q)} e^{i z (p_2+p_2^\prime+q)}
e^{i y (p_1-p_2)} \pr{ \frac{1}{\d^2} e^{i y^{\prime\prime} (p_1^\prime-p_2^\prime)} }
\nn \\
& ~~~~\times \frac{
 i^{a+b+a^\prime+b^\prime+c} 
}{
\Gamma(a)\Gamma(b)\Gamma(a^\prime)\Gamma(b^\prime)
\Gamma(c)}
p_1^{a-1} p_2^{b-1} {p_1^\prime}^{a^\prime-1} {p_2^\prime}^{b^\prime-1} q^{c-1}~
\theta (p_1) \theta (p_2) \theta (p_1^\prime) \theta (p_2^\prime) \theta (q) ~,
\end{align}
where the momenta $p_i$ and $p_i^\prime$ are the momenta of active fermions and $q$ comes from the spectators. 
We take the standard Fourier transformation to obtain the momentum space matrix element
\begin{align}
&\int e^{i P x - i P^\prime z} dxdydz 
\pr{\frac{\d}{\d y^{\prime\prime}}}^{-2} G(x, y, y^{\prime\prime}, z) 
\nn \\
&=~ (2\pi) \delta(P-P^\prime) A_{\kvec,\kvec^\prime} \int dp_1 dp_2 dp_1^\prime dp_2^\prime dq\, \nn \\
&~~~~ \times (2\pi) \delta( p_1+p_1^\prime+q - P ) (2\pi) \delta( p_2+p_2^\prime+q - P ) \nn \\
&~~~~ \times \frac{i^{a+b+a^\prime+b^\prime+c} 
}{
\Gamma(a)\Gamma(b)\Gamma(a^\prime)\Gamma(b^\prime)
\Gamma(c)}  \times \frac{
-p_1^{a-1} p_2^{b-1} {p_1^\prime}^{a^\prime-1} {p_2^\prime}^{b^\prime-1} q^{c-1}
}{(p_1^\prime-p_2^\prime)^2} \nn \\
&\equiv~ 
(2\pi) \delta(P-P^\prime)\frac{
    4\pi^2 i^{\Delta+\Delta^\prime-2}
    P^{\Delta+\Delta^\prime}
    A_{\kvec,\kvec^\prime}
}{\Gamma(\Delta+\Delta^\prime-1)} I(a,b,a^\prime,b^\prime)~,
\end{align}
where the momentum on the denominator comes from acting the $1/\d^2$ on the exponential of $y^{\prime\prime}$. The spacial integral becomes momentum conservation. As usual, it can normalize the total external momentum, and express the integral in terms of the momentum fraction.
A particularly convenient substitution is
\begin{align}
p_1 &\equiv x_1 x_2 ~,\nn \\
p_2 &\equiv x_1 x_3~, \nn \\
q &\equiv 1-x_1 ~,\nn \\
p_1^\prime &= x_1(1-x_2)~, \nn \\
p_2^\prime &= x_1(1-x_3) ~,
\end{align}
which separates the active part and the spectators.
We have thus worked out a general formula to evaluate the gauge interaction matrix elements, term by term from Wick contraction, as a momentum integral
\begin{align}\label{eq:gauge-interaction-momentum}
I(a,b,a^\prime,b^\prime) =& \frac{
\Gamma(\Delta+\Delta^\prime-1)
}
{
\Gamma(a)\Gamma(b)\Gamma(a^\prime)\Gamma(b^\prime)
\Gamma(c)}
\int \, dx_1 x_1^{a+b+a^\prime+b^\prime-4}(1-x_1)^{c-1} \nn \\
&~~\times 
\int dx_2 dx_3 \, 
\frac{x_2^{a-1} (1-x_2)^{a^\prime-1} x_3^{b-1} (1-x_3)^{b^\prime-1} }{(x_2-x_3)^2} ~.
\end{align}
It is nice that the momentum fraction of the spectators factors out of the principal value integral, making it possible to find a closed-form expression for the active part. The integral over spectators' momentum is
\begin{align}
\int \, dx_1 x_1^{a+b+a^\prime+b^\prime-4}(1-x_1)^{c-1}
= \frac{\Gamma (c) \Gamma \left(a+b+a'+b'-3\right)}{\Gamma \left(a+b+c+a'+b'-3\right)} ~.
\end{align}

Now we are left with the integral of $x_2$ and $x_3$
\begin{align}
I(a,b,a^\prime,b^\prime) &= 
\frac{\Gamma \left(a+b+a'+b'-3\right) }{
    \Gamma(a)\Gamma(a^\prime)\Gamma(b)\Gamma(b^\prime)
}
\int dx_2 dx_3 \, 
\frac{x_2^{a-1} (1-x_2)^{a^\prime-1} x_3^{b-1} (1-x_3)^{b^\prime-1} }{(x_2-x_3)^2}
\nn \\
&= \frac{\Gamma \left(a+b+a'+b'-3\right) }{
    \Gamma(a)\Gamma(a^\prime)\Gamma(b)\Gamma(b^\prime)
} \times \sum_{m_1}^{a^\prime-1}  \sum_{m_2}^{b^\prime-1}
\binom{a^\prime-1}{m_1} \binom{b^\prime-1}{m_2} 
I_1(a+m_1-1,b+m_2-1)~,
\end{align}
where the integral is reduced to the most general elemental form
\begin{align}\label{eq:gauge-master-integral-1}
I_1(a,b) &\equiv 
\int dx_2 dx_3 \, \frac{
    x_2^a x_3^b 
}{(x_2-x_3)^2} + (\text{self-energy shift}) ~.
\end{align}
The integral is divergent. The divergence cancels out with a self-energy term from normal ordering the interaction term. With the self-energy term, the result becomes
\begin{align}
I_1(a,b)&= \text{PV} \int dx_2 dx_3 \, \frac{
        x_2^a x_3^b - x_2^{a+b}
    }{(x_2-x_3)^2} \nn \\
   &= \frac{a H_a+b H_b-1}{a+b} - H_{a+b-1}~,
\end{align}
where PV stands for taking the principal value, following the principal value prescription 
\begin{equation}
\text{PV} \int \frac{\psi(k)dk}{k^2} \equiv \frac{1}{2} \int 
\frac{\psi(k+i \epsilon)dk}{(k+i\epsilon)^2} +
\frac{1}{2} \int \frac{\psi(k-i \epsilon)dk}{(k-i\epsilon)^2} ~.
\label{eq:PVdef}
\end{equation}

\subsection{\texorpdfstring{$a = 0$}{a=0} case}
In (\ref{eq:gauge-5pt-func}) one or more of the four numbers ($a$, $b$, $a^\prime$ or $b^\prime$) can vanish. Without loss of generality, we can set $a=0$, and we have the freedom to integrate with respect to either $y$ or $y^{\prime\prime}$  and get the same answer. We can just compute
\begin{align}\label{eq:gauge-5pt-trivial}
&\int^y dy^{\prime} \int^{y^\prime} dy^{\prime\prime} G(x, y, y^{\prime\prime}, z) 
\nn \\
&\equiv 
A_{\kvec,\kvec^\prime}
\times
\left(
    \int^y dy^{\prime} \int^{y^\prime} dy^{\prime\prime}
    \frac{1}{(y^{\prime\prime}-z)^b}
\right)
\frac{1}{(x-y)^{a^\prime}}
\frac{1}{(y-z)^{b^\prime}}
\frac{1}{(x-z)^{c}} ~.
\end{align}
The integral over $y^{\prime\prime}$ is just the naive indefinite integral because the boundary value vanishes, except for $b=2$ case,
\begin{align}\label{eq:gauge-interaction-momentum-trivial}
\int^y dy^{\prime} \int^{y^\prime} dy^{\prime\prime}\frac{1}{(y-z)^b} = \begin{cases}
\frac{1}{(b-1)(b-2)(y-z)^{b-2}} & b>2\\
\text{depends on the boundary condition} & b=2
\end{cases} ~.
\end{align}
Unless $b=2$, we can be sloppy about the boundary condition of the integral because the boundary value at $y^{\prime\prime}\rightarrow \infty$ and $y^{\prime}\rightarrow \infty$ vanish, and the integral over $y^{\prime\prime}$ is just the naive indefinite integral. The Fourier transform is known, using a general integral identity 
\begin{equation}
\int \! dx \, dy \, dz \frac{e^{i(px-p'z)}}{(x-y)^A (y-z)^B (x-z)^C} \! = \! (2\pi)\delta(p-p') \frac{4\pi^2\Gamma(A+B-1) p^{A+B+C-2}}{\Gamma(A)\Gamma(B)\Gamma(A+B+C-1)}~,
\end{equation}
we obtain the final result for $b>2$
\begin{align}
I(0,b,a^\prime,b^\prime) 
=\,& \frac{\Gamma \left(b+a^\prime+b^\prime-3\right)}{(b-2) (b-1) \Gamma \left(a^\prime\right) \Gamma
   \left(b+b^\prime-2\right)}~.
\end{align}

\subsection{\texorpdfstring{$a = 0$}{a=0}, \texorpdfstring{$b = 2$}{b=2} case}
The $b\rightarrow2$ limit of (\ref{eq:gauge-interaction-momentum-trivial}) depends on the boundary condition, so we proceed in the momentum space. Like the $t$-channel case, we write the spacial factors in (\ref{eq:gauge-5pt-trivial}) in the momentum space and Fourier transform the overall formula with respect of $x$ and $z$. Note that $p_1$ is missing since $(x-y)$ factor is missing. 
\begin{align}\label{eq:gauge-s-channel-special-case}
& \int e^{i P x - i P^\prime z} dxdydz
\int^y dy^{\prime} \int^{y^\prime} dy^{\prime\prime}
G(x, y, y^{\prime\prime}, z)  \nn \\
=&~ (2\pi) \delta(P-P^\prime) A_{\kvec,\kvec^\prime} 
\int dp_2 dp_1^\prime dp_2^\prime (2\pi)\delta(p_1^\prime+q-P) 
\times (2\pi) \delta(p_2+p_2^\prime +q -P) \nn \\
&~~\times \frac{i^{a^\prime+b^\prime+c+2}}{\Gamma(b)\Gamma(c)\Gamma(a^\prime)\Gamma(b^\prime)}
\times \frac{(-1) p_2 (p_1^\prime)^{a^\prime-1}(p_2^\prime)^{b^\prime-1} q^{c-1}} 
{(p_1^\prime - p_2^\prime)^2}
 \nn \\
=&~ (2\pi) \delta(P-P^\prime)\frac{
    4\pi^2 i^{\Delta+\Delta^\prime-2}
    P^{\Delta+\Delta^\prime}
    A_{\kvec,\kvec^\prime}
}{\Gamma(\Delta+\Delta^\prime-1)} I (0,2,a^\prime,b^\prime,c)~,
\end{align}
where we used the momentum conservation 
\begin{equation}
 p_2 = p_1^\prime - p_2^\prime~.
\end{equation}
We have thus defined the matrix elements as momentum space integrals. We can further parameterize the momenta as
\begin{align}
p_2 &\equiv x_1 (1-x_2)~, \nn \\
q &\equiv 1-x_1~, \nn \\
p_1^\prime &= x_1~, \nn \\
p_2^\prime &= x_1 x_2 ~,
\end{align}
and try working out the integral.
\begin{align}
I(0,2,a^\prime,b^\prime,c) = \frac{
    \Gamma(\Delta+\Delta^\prime-1)
}{\Gamma(b)\Gamma(c)\Gamma(a^\prime)\Gamma(b^\prime)} 
\int dx_1 \, x_1^{a^\prime +b^\prime -2} (1-x)^{c-1}
\int dx_2 \frac{x_2^{b^\prime-1}}{1-x_2}~.
\end{align}
The $x_1$ integral is finite,
\begin{align}
\int dx_1 \, x_1^{a^\prime +b^\prime -2} (1-x)^{c-1}
= \frac{\Gamma(a^\prime+b^\prime-1)\Gamma(c)}
   {\Gamma(a^\prime+b^\prime+c-1)}~.
\end{align}
The other integral is divergent. We need to find the scheme for the self-energy regulator. The key is that the wave functions that contract with $\frac{1}{\d^2}\psi^\dagger\psi$ needs to be anti-symmetrized under $\psi^\dagger \leftrightarrow \psi$, i.e. under $p_1^\prime \leftrightarrow p_2^\prime$. Thus the correct self-energy shift is 
\begin{align}
\frac{ (p_1^\prime)^{a^\prime-1}(p_2^\prime)^{b^\prime-1} }
{(p_1^\prime - p_2^\prime)^2} &\rightarrow
\frac{ (p_1^\prime)^{a^\prime-1}(p_2^\prime)^{b^\prime-1} 
 - (p_2^\prime)^{a^\prime+b^\prime-2}
}
{(p_1^\prime - p_2^\prime)^2} \nn \\
\int dx_2 \frac{x_2^{b^\prime-1}}{1-x_2}
&\rightarrow \text{PV} \int dx_2 \frac{
    x_2^{b^\prime-1} - x_2^{a^\prime+b^\prime-2}
}{1-x_2}
= - H_{b-1} + H_{a+b-2}~.
\end{align}
We finally have
\begin{align}
I(0,2,a^\prime,b^\prime) 
=\,& \frac{\Gamma \left(a^\prime+b^\prime-1\right)}{\Gamma
   \left(a^\prime\right) \Gamma \left(b^\prime\right)}
   \left(H_{a^\prime+b^\prime-2}-H_{b^\prime-1}\right) ~.
\end{align}

\bibliographystyle{utphys}
\bibliography{citations}
\end{document}